\documentclass[aps,twocolumn,showpacs,amsmath,amssymb,floatfix]{revtex4}
\usepackage{young}
\usepackage{graphicx}
\usepackage{color}
\usepackage[latin1]{inputenc} 
\usepackage{young}

\def\nn{\nonumber}

\def \bc {\begin{center}}
\def \ec {\end{center}}
\def \bi {\begin{itemize}}
\def \ei {\end{itemize}}

\def \ba {\begin{array}}
\def \ea {\end{array}}

\def \bea {\begin{eqnarray}}
\def \eea {\end{eqnarray}}

\def \be {\begin{equation}}
\def \ee {\end{equation}}

\def \um {\frac{1}{2}}
\def\tr {\mathrm{tr}}
\def\cD {{\cal D}}

\newcommand{\la}{\langle}
\newcommand{\ra}{\rangle}

\begin{document}

\title{Interlayer coherence and entanglement in bilayer quantum Hall states at filling factor $\nu=2/\lambda$}

\author{M. Calixto and E. P\'erez-Romero}

\affiliation{Departamento de Matem\'atica Aplicada, Universidad de Granada,
Fuentenueva s/n, 18071 Granada, Spain}


\begin{abstract}
We study coherence and entanglement properties of the state space of a composite bi-fermion (two electrons pierced by $\lambda$ magnetic flux lines) 
at one Landau site of a bilayer quantum Hall system. In particular, interlayer imbalance and entanglement (and its fluctuations) are analyzed 
for a set of $U(4)$ coherent (\emph{quasiclassical}) states generalizing the standard pseudospin $U(2)$ coherent states for the spin-frozen case. 
The interplay between spin and pseudospin degrees of freedom opens new possibilities with regard to the spin-frozen case. 
Actually, spin degrees of freedom make interlayer entanglement more effective and robust under perturbations than in the spin-frozen situation, 
mainly for a large number of flux quanta $\lambda$. Interlayer entanglement of an equilibrium thermal state and its dependence with temperature 
and bias voltage is also studied for a pseudo-Zeeman interaction.
\end{abstract} 


\maketitle

\section{Introduction}

Quantum Hall Effect (QHE) keeps catching researchers' attention owing to its peculiar features mainly related to 
quantum coherence and the emergence of a new class of particles called ``composite fermions'', due to collective behavior shared with 
superconductivity and Bose-Einstein condensation phenomena. In fact, the physics of the bilayer quantum Hall (BLQH) systems, 
made by trapping electrons in two thin layers 
at the interface of semiconductors, is quite rich owing to unique effects originating in the intralayer and interlayer 
coherence developed by the interplay between the spin and the layer (pseudospin) degrees of freedom.
For example, the presence of interlayer coherence in bilayer quantum Hall states has been examined 
by magnetotransport experiments \cite{Interlayercoh}, where electrons are transferable between the two layers by applying bias voltages 
and the interlayer phase difference is tuned by tilting the sample. Also,  
anomalous (Josephson-like)  tunneling current between the two layers at zero bias voltage were predicted in Refs. \cite{EzawafracBLQH,WenZee,EisenMac}, 
whose first experimental indication was obtained in Ref. \cite{SpiEisPWest}. Other original studies on spontaneous interlayer coherence in 
BLQH systems are \cite{KunYang1,KunYang2}.

Spin and pseudospin quantum degrees of freedom are 
correlated in BLQH systems and entanglement properties have also been studied in, for example, 
Refs. \cite{Doucot,Schliemann,Yusuke}, mainly at filling factor $\nu=1$. An appropriate description of quantum correlations is of great relevance in
quantum computation and information theory, a field which has also attracted a huge degree of attention. Actually, one can find 
quantum computation proposals using BLQH systems in, for example, \cite{Scarola,Yang,Park}. In this article we 
also address the interesting problem of quantum coherence and entanglement in BLQH systems at fractions of $\nu=2$ (perhaps a less known case), in the hope 
that our theoretical considerations contribute to eventually implement feasible large
scale quantum computing in BLQH systems by  engineering quantum Hall states. For this purpose, controllable entanglement, 
robustness of qubits (long decoherence time) and ease qubit measurement 
are crucial. Concerning qubit measurement (and general reconstruction of quantum states), coherent states (which are often said to be the most 
classical of all states of a dynamical quantum system) have been widely used to reconstruct the quantum state of light \cite{Leonhardt},  
pure spin sates \cite{Amiet,Brif}, etc, by using tomographic, spectroscopic, interferometric, etc, techniques. The existence of 
interlayer and intralayer coherence in BLQH systems has also been evidenced (as commented in the previous paragraph), and 
we think that is its worth studying coherent states (CS in the following) for the ``Grassmannian'' $\nu=2/\lambda$ case, which is perhaps less known than 
the ``complex projective'' (totally symmetric) $\nu=1/\lambda$ case. The subject of CS is not only important for the quantum state reconstruction 
problem, but also to analyze the phase diagram of Hamiltonian models undergoing a quantum phase transition (like the well studied spin-ferromagnet
and pseudo-spin-ferromagnet phases at $\nu=1$). This is the spirit of Gilmore's algorithm \cite{Gilmore2}, which makes use of CS 
as variational states to approximate the ground state energy, to study the classical, thermodynamic or mean-field, limit of some critical quantum models 
and their phase diagrams. For the BLQH system at $\nu=2$, the variational ground state energy per Landau site (with a Hamiltonian 
consisting on Coulomb plus Zeeman-pseudo-Zeeman terms) has been analyzed (see e.g.  \cite{EzawaBook}). 
The variational, $SU(4)$-invariant, ground state is a homogeneous (coherent) 
state parametrized by eight independent variables [or four complex parameters $z^\mu, \mu=0,1,2,3$, in our notation; see later on equations 
(\ref{cs1},\ref{cs11})] which are related to the eight so-called Goldston modes in the $SU(4)$-invariant system. Minimizing the 
variational ground state energy within this parameter space (eight-dimensional energy surface) reveals the existence of three phases at 
$\nu=2$: spin, ppin and canted phases (see e.g. \cite{EzawaBook}) as we move through the 
BLQH Hamiltonian coupling constants (bias voltage, tunneling, Zeeman strength, etc). At the minimum, the eight coherent (ground) state 
parameters now depend on these Hamiltonian coupling constants and we could manipulate them to generate not only specific CS but also interesting combinations like 
the so-called Schr\"odinger cat states. The existence of Schr\"odinger cat states has been evidenced in other physical models undergoing 
a quantum phase transition like the Dicke model for atom-radiation interaction 
(see e.g. \cite{casta2,epl2012,renyipra,husidi}) and vibron models for molecules 
\cite{curro,husivi,entangvib,entangvib2}, among others. In this paper we provide explicit expressions 
of CS for BLQH at general fractions of $\nu=2$ and we study some physical properties like 
interlayer imbalance and entanglement and their fluctuations. We believe that the CS discussed in this paper will also be 
of importance when studying many other BLQH issues like the aforementioned phase diagrams.

In the BLQH system, one Landau site can 
accommodate four isospin states $|b\uparrow\rangle, |b\downarrow\rangle, |a\uparrow\rangle$ and $|a\downarrow\rangle$ in
the lowest Landau level,  where $|b\uparrow\rangle$ (resp. $|a\downarrow\rangle$) means  that the
electron is in the bottom layer ``$b$'' (resp. top  layer ``$a$'') and its spin is up (resp. down), and so on.
Therefore, the underlying group structure in each Landau level of the BLQH system is enlarged from spin symmetry $U(2)$ to 
isospin symmetry $U(4)$. The driving force of quantum coherence is the Coulomb exchange interaction, which is described by an anisotropic $U(4)$
nonlinear sigma model  in BLQH systems \cite{Interexchange,EzawaBook}. Actually, it is the interlayer exchange interaction which
develops the interlayer coherence. The lightest topological charged excitation in the BLQH system is a (complex projective)
$\mathbb CP^3=U(4)/[U(1)\times U(3)]$ skyrmion for filling factor $\nu=1$ and a (complex Grassmannian)
$\mathbb G_2^4=U(4)/[U(2)\times U(2)]$ bi-skyrmion 
for filling factor $\nu=2$ (see \cite{KunYang3} for similar studies in graphene and the charge of these excitations). 
The Coulomb exchange interaction for this last case is described by a Grassmannian
$\mathbb G_2$ (from now on, we omit the superscript in $\mathbb G_2^4$)  sigma model and the 
dynamical field  is a  Grassmannian field $Z=z^\mu\sigma_\mu$ \cite{Ezawabisky} ($\sigma_\mu$ denote Pauli matrices plus identity) 
carrying four complex field degrees of freedom 
$z^\mu\in\mathbb C$, $\mu=0,1,2,3$ (see later on Sec. \ref{sec2}). As commented before, the parameter
space characterizing the $SU(4)$-invariant ground state in the BLQH system at $\nu=2$ is precisely $\mathbb G_2$ \cite{EzawaPRB}.

We would also like to say that higher-dimensional generalizations of the Haldane's sphere picture \cite{Haldane} of FQHE appeared after 
Zhang and Hu four-dimensional generalization of QHE \cite{Zhang}. Just to mention several studies of QHE on general manifolds and 
their CS like for example: 
torus $\mathbb T^2$ \cite{frachall}, complex projective $U(N)/[U(N-1)\times U(1)]$ \cite{Karabali}, Bergman ball 
$U(N,1)/[U(N)\times U(1)]$ \cite{Jellal1}, flag manifold $U(3)/U(1)^3$ \cite{Jellal2}, and many others. Coherent states 
have also been worked out in these higher-dimensional generalizations, which help to perform a semi-classical 
analysis and to construct effective Wess-Zumino-Witten 
actions for the edge states. Similar contructions could also be done for BLQH systems at $\nu=2/\lambda$ with the help 
of the Grassmannian CS that we discuss  in this paper.

Two electrons in one Landau site must form an antisymmetric state due to Pauli exclusion principle and this leads to a 6-dimensional 
irreducible representation of $SU(4)$, which is usually divided into spin and pseudospin sectors. 
The composite-fermion field theory \cite{JainPRL,Jainbook} and experiments reveal the existence of new fractional QH states in the bilayer system 
\cite{EzawafracBLQH}. For fractional values of the filling factor, $\nu=2/\lambda$, the composite fermion interpretation is that of 
two electrons pierced by $\lambda$ magnetic flux lines. The mathematical structure of the $d_\lambda$-dimensional \eqref{dimension} Hilbert space 
${\cal H}_\lambda(\mathbb G_2)$ for two composite particles in one Landau site has been studied in a recent article 
\cite{GrassCSBLQH} by us, where we have also constructed the set of CS labeled by points of $\mathbb G_2$. 
For $\lambda$ odd, wave functions turn out to be antisymmetric (composite fermion) and for $\lambda$ even, wave functions are symmetric (composite 
boson), see later on eq. \eqref{statistics}. Now we want to analyze some physical properties of these ``quasi-classical'' states, like interlayer imbalance, 
entanglement and their fluctuations, comparing them 
with the simpler spin-frozen case, to evaluate the effect played by spin and extra $U(4)$ isospin operators. In particular, we observe that 
the number $\lambda$ of flux lines, for filling factor $\nu=2/\lambda$, affects non-trivially the interlayer entanglement of CS.

The paper is organized as follows. In Section \ref{secfrozen} we start by briefly analyzing the easier spin-frozen case 
and considering only the $U(2)$ pseudospin structure; the $U(2)$ pseudospin-$s$ operators, Hilbert space and CS  
are discussed in a oscillator (bosonic) realization related to magnetic flux quanta attached to the electron in the fractional filling 
factor case. This oscillator construction somehow reminds the quasi-spin formalism introduced in the two-mode approximation of 
Bose-Einstein condensates in a double-well 
potential, with the role of the two wells played now by the two layers. 
We compute interlayer imbalance and entanglement of pseudospin-$s$  CS to 
later appreciate the similitudes and differences between the spin-frozen and the more involved isospin-$\lambda$ $U(4)$ case. 
Section \ref{sec2} is devoted to a brief exposition of the operators (spin, pseudospin, etc.) 
and the structure of the Hilbert space 
${\cal H}_\lambda(\mathbb G_2)$ of two electrons, at one Landau site of the lowest Landau level, pierced by $\lambda$ 
magnetic flux lines. For this purpose,  we introduce
an oscillator realization of the $U(4)$ Lie algebra in terms of eight boson creation, $a_\mu^\dag, b_\mu^\dag$, and annihilation,
$a_\mu, b_\mu, \mu=0,1,2,3$, operators. Then an orthonormal basis of ${\cal H}_\lambda(\mathbb G_2)$, in terms of Fock 
states, is explicitly constructed, and a set of CS  $\{|Z\ra\}$,  labeled by points 
$Z\in\mathbb G_2$, is built  as definite superpositions of the basis states which remind Bose-Einstein condensates. 
 The simpler (lower-dimensional) case $\lambda=1$ is explicitly written, leaving the more involved (higher-dimensional) $\lambda>1$ case for Appendices 
\ref{basissubsec} and \ref{subsecCS}  (the reader can find much more 
information about the mathematical structure of the state space ${\cal H}_\lambda(\mathbb G_2)$ in \cite{GrassCSBLQH}). In Section  
\ref{secimb} we analyze interlayer imbalance and its fluctuations in a general Grassmannian $U(4)/U(2)^2$, isospin-$\lambda$, CS 
$|Z\ra$, which generalizes the interlayer imbalance of a pseudospin-$s$ CS $|z\rangle$, recovering the spin-frozen situation as a particular case.  
In Section \ref{entangsec} we examine the interesting problem of interlayer entanglement for basis states and CS, 
accessed through the calculation of the \emph{purity} (and its fluctuations), linear and Von Neumann entropies 
of the reduced density matrix to one of the layers. We 
find out that spin degrees of freedom play a role in the interlayer entanglement by, for example,
making it more robust than in the spin-frozen case. Interlayer entanglement of an equilibrium thermal (mixed) state and its 
dependence with temperature and bias voltage is also studied for a pseudo-Zeeman interaction. 
Section \ref{comments} is devoted to conclusions and outlook.

\section{The simpler U(2) spin-frozen case}\label{secfrozen}

We shall start by briefly analyzing the spin-frozen case and considering only the $U(2)$ pseudospin 
structure by assigning up and down pseudospins to the electron on the top $a$ and bottom $b$ layers, 
respectively (see e.g. \cite{EzawaBook} for a standard reference on this subject). This approximation 
is valid when the Zeeman
energy is very large and all spins are frozen into their polarized states. We shall 
recover the spin degree of freedom in Section \ref{sec2}. The electron configuration is 
described by the total number density $\rho$ and the pseudospin density 
$\vec{\mathcal P}=(\mathcal P_1,\mathcal P_2,\mathcal P_3)$, whose direction is controlled 
by applying bias voltages which transfer electrons between the two layers. We shall restrict 
ourselves in this Section to one electron at one Landau site of the lowest Landau level, pierced by $2s$ magnetic flux lines, with $s$ the pseudospin. 
The operators $a^\dag$ and $b^\dag$ (resp. $a$ and $b$) create (resp. annihilate) flux quanta attached to 
the electron at the top and bottom layers, respectively. If we denote by $\mathcal Z=\begin{pmatrix} a\\  b
\end{pmatrix}$ and $\mathcal Z^\dag=(a^\dag,b^\dag)$ the two-component electron ``field'' and its 
conjugate, then the pseudospin density operator can be compactly written as
\be\mathcal{P}_\mu=\um \mathcal Z^\dag\sigma_\mu \mathcal Z, \;\; \mu=0,1,2,3,\label{spinfrozen}\ee
where $\sigma_\mu, \mu=1,2,3,$ denote the usual three Pauli matrices and 
$\sigma_0$ is the $2\times 2$ identity matrix. The representation \eqref{spinfrozen} 
resembles the usual  Jordan-Schwinger boson realization  for spin. Note that 
$2\mathcal{P}_0=\mathcal Z^\dag\mathcal Z=a^\dag a+b^\dag b$ represents the total number $N$ 
of flux quanta, which is fixed to $N=2s$ with $s$ the pseudospin. The pseudospin third 
component $\mathcal P_3=\um(a^\dag a-b^\dag b)$ measures the population imbalance between 
the two layers, whereas $\mathcal P_\pm=\mathcal P_1\pm\mathcal P_2$ are tunneling (ladder) operators 
that transfer quanta from one layer to the other and create interlayer coherence [see later on eq. \eqref{su2cs}].  
The boson realization \eqref{spinfrozen} defines a unitary representation of the pseudospin $U(2)$ 
operators $\mathcal P_\mu$ on the Fock space expanded by the orthonormal basis states 
\be
|n_a\rangle\otimes|n_b\rangle=\frac{(a^\dag)^{n_a}(b^\dag)^{n_b}}{\sqrt{n_a!n_b!}}|0\ra, \label{bosrepreu2}
\ee
where $|0\ra$ denotes the Fock vacuum and $n_\ell$ the occupancy numbers of layers $\ell=a,b$. 
The fact that the total number of quanta is constrained to  $n_a+n_b=2s$ indicates that the representation
\eqref{spinfrozen} is reducible in Fock space. A $(2s+1)$-dimensional irreducible (Hilbert) subspace 
$\mathcal H_s(\mathbb S^2)$ carrying a unitary representation of $U(2)$ with pseudospin $s$ is expanded 
by the $\mathcal P_3$ eigenvectors
\be
|k\ra\equiv|s+k\rangle_a\otimes|s-k\rangle_b=
\frac{\varphi_k(a^\dag)}{\sqrt{\frac{(2s)!}{(s+k)!}}}
\frac{\varphi_{-k}(b^\dag)}{\sqrt{\frac{(2s)!}{(s-k)!}}}|0\ra,\label{basisinfocksu2}
\ee
with $|0\ra$ the Fock vacuum and $k=-s,\dots,s$ the corresponding eigenvalue (pseudospin third component). We have made use of the monomials  
$\varphi_k(z)=\binom{2s}{s+k}^{1/2}z^{s+k}$ as a useful notation to generalize the Fock space representation 
\eqref{basisinfocksu2} of the pseudospin $U(2)$ states $|k\rangle$, to the 
isospin $U(4)$ states $|{}{}_{q_a,q_b}^{j,m}\ra$ in eq. \eqref{basisvec}, explicitly written  later in eq. \eqref{basisinfock2}. 
This construction resembles Haldane's sphere picture \cite{Haldane} for spinning monolayer QH systems, 
where $s$ is also related to the ``monopole strength'' in the sphere $\mathbb S^2$. 

As already said, tunneling between the two layers creates interlayer coherence, which can be described by 
pseudospin-$s$ CS
\be
|z\rangle=\frac{e^{z \mathcal{P}_+}|-s\rangle}{(1+|z|^2)^{s}}=
\frac{\sum_{k=-s}^s\varphi_k(z)|k\rangle}{(1+|z|^2)^{s}},\label{su2cs}
\ee
obtained as an exponential action of the tunneling (rising) operator $\mathcal{P}_+$ on the lowest-weight state $|-s\rangle$ (all 
quanta at the bottom layer $b$) with tuneling strength $z$, a complex number usually parametrized as 
$z=\tan(\theta/2)e^{i\phi}$, related to the stereographic projection of a 
point $(\theta,\phi)$ (polar and azimutal angles) of the Bloch sphere $\mathbb S^2=U(2)/U(1)^2$ onto the complex plane. 
The modulus and phase of $z$ have a BLQH physical meaning which will be explained in the next Subsection. From the mathematical point of view, 
pseudospin-$s$ CS are normalized (but not orthogonal), as can be seen from the CS overlap
\be\langle z'|z\rangle=\frac{(1+\bar{z}'z)^{2s}}{(1+|z'|^2)^s(1+|z|^2)^s},\label{su2overlap}\ee
and they constitute an overcomplete set fulfilling the resolution of the identity $1=\int_{{\mathbb S^2}}|z\rangle\langle z|d\mu(z,\bar z)$, 
with $ d\mu(z,\bar z)=\frac{2s+1}{\pi}\sin\theta d\theta d\phi$ the solid angle.

Pseudospin-$s$ CS  also accurately describe the physical properties of many macroscopic quantum systems like Bose-Einstein condensates 
in a double-well potential, two-level systems, superconductors, superfluids, etc. A more familiar Fock-space representation of pseudospin-$s$ 
CS [equivalent to \eqref{su2cs}] 
as a two-mode Bose-Einstein condensate is given by ($|0\rangle$ denotes the Fock vacuum)
\be
|z\rangle=\frac{1}{\sqrt{(2s)!}}\left(\frac{b^\dag+za^\dag}{\sqrt{1+|z|^2}}\right)^{2s}|0\ra.\label{su2BE}
\ee
In this context, the polar angle $\theta$ is related to the population imbalance and the azimuthal angle $\phi$ is the relative phase of the 
two spatially separated Bose-Einstein condensates. 
Both quantities can be experimentally determined in terms of matter wave interference experiments as it is shown in Refs. 
\cite{prl92,prl95,science307}. Let us see how both quantities also describe imbalance population and phase coherence between layers 
in the spin-frozen BLQH system.

\subsection{Interlayer coherence and imbalance fluctuations}\label{imbsubsec}

Standard (harmonic oscillator) CS exhibit Poissonian number statistics for the probability
of finding $n$ bosons, so that standard deviation $\Delta n$ is large. These large
fluctuations  of the occupation number are typical in superfluid phases. Here we shall compute the mean
value $\la\mathcal{P}_3\rangle$ and standard deviation $\Delta \mathcal{P}_3$ of the interlayer imbalance operator
$\mathcal{P}_3$ in a pseudospin-$s$ CS $|z\rangle$. Taking into account
that the pseudospin-$s$ basis states $\{|k\ra, k=-s,\dots,s\}$ are eigenstates of $\mathcal{P}_3$, namely $\mathcal{P}_3|k\ra=k|k\ra$, one can
easily compute the (spin-frozen) imbalance $\iota$ and its fluctuations  $\Delta\iota$ per flux in a pseudospin-$s$  CS \eqref{su2cs}  as
\bea
\iota&=&\frac{\la z|\mathcal{P}_z|z\ra}{s}=\frac{\la \mathcal{P}_z\ra}{s}=\frac{|z|^2-1}{|z|^2+1}=-\cos\theta,\label{imbsu2}\\ 
\Delta\iota&=&\sqrt{\frac{\langle  \mathcal{P}_z^2\rangle-\langle \mathcal{P}_z\rangle^2}{s}}=
\frac{\sqrt{2}|z|}{1+|z|^2}=\frac{\sin\theta}{\sqrt{2}}.\nn
\eea
\begin{figure}
\includegraphics[width=8cm]{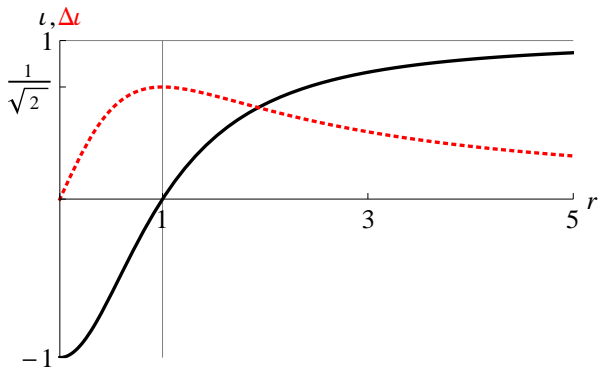}
\caption{ \label{imbalance} (Color online)  Imbalance $\iota$  (black line)
and its standard deviation $\Delta \iota$ (dotted red line) per flux as a function of the CS parameter $r=|z|$.}
\end{figure}
In Figure \ref{imbalance} we see that the imbalance $\iota$ is $-1$ at $r=|z|=0$, for which
the CS is $|z_0\rangle=|-s\rangle$ (the lowest-weight state), that is, all quanta at the bottom layer $b$. The imbalance
$\iota$ is 0 at $r=1$ (the
balanced case) and $\iota\to 1$ when $r\to\infty$, for which the CS is  $|z_\infty\rangle=|s\rangle$ (the highest-weight state),
that is, all quanta at the top layer $a$. The standard deviation $\Delta \iota$ is maximum at $r=1$, with
$\Delta \iota(1)=1/\sqrt{2}$, and tends to zero at $r=0$ and when $r\to\infty$. This indicates that
the largest fluctuations occur at $r=1$ ($\theta=\pi/2$). Note that both $\iota$ and $\Delta \iota$ are invariant under inversion $r\to 1/r$, namely
$\iota(r)=-\iota({1}/{r})$ and $\Delta \iota(r)=\Delta \iota({1}/{r})$, the point  $r=1$ being a fixed point. 
The other interesting physical magnitude is the interlayer phase difference $\phi=\arctan\la\mathcal{P}_y\ra/\la\mathcal{P}_x\ra$, which 
was evidenced in  \cite{Eisensteinphase} for BLQH systems. A robust interlayer phase difference is essential to design BLQH quantum bits 
\cite{Yang} which could enable large-scale quantum computation \cite{Scarola,Park}.

The spinning case will provide more degrees of freedom than the spin-frozen case to play with,
since we will have extra isospin operators in $u(4)$ to create interlayer coherence (see later on Sec. \ref{secimb}). 

\subsection{Interlayer entanglement}\label{entsubsec}

In the pseudospin state space $\mathcal{H}_s(\mathbb S^2)$, we shall consider the bipartite quantum system given by layers $a$ and $b$.
At first glance, the basis states \eqref{basisinfocksu2} are a direct product and do not entangle both layers. However, we shall see in Sec. 
\ref{entangsec} that the introduction of spin creates new quantum correlations on the basis states. On the contrary, 
pseudospin-$s$ CS do entangle layers $a$ and $b$. Indeed, considering the density matrix $\varrho=|z\rangle\langle z|$ 
and the expression of the pseudospin-$s$ basis states 
$|k\rangle$ as a direct product of Fock states \eqref{basisinfocksu2} in layers $a$ and $b$, 
the reduced density matrix (RDM) to layer $b$ is 
\be
\varrho_b=\tr_a(\varrho)=\sum_{n=0}^{2s}\gamma_n(r)|2s-n\ra_b\la 2s-n|, \label{redsu2}
\ee
which turns out to be diagonal with eigenvalues $\gamma_n(r)=\binom{2s}{n}{r^{2n}}/{(1+r^2)^{2s}}$ a function of 
$r=|z|$. This expression coincides with the result of Ref. \cite{annalcasta} for entanglement of spin CS arising
in two-mode ($a$ and $b$) Bose-Einstein condensates; see \cite{entangsu2su11} for other results on entangled $SU(2)$ CS 
and \cite{entangcoh-review} for a review on this subject. The purity $p_s=\tr(\varrho_b^2)$ of \eqref{redsu2} 
as a function of $r$ is then
\be
p_s(r)=\sum_{n=0}^{2s}\gamma_n^2(r)=\frac{\sum_{n=0}^{2s} \binom{2s}{n}^2 r^{4n}}{(1+r^2)^{4s}}.\label{puritysu2eq}\ee
This function is also inversion invariant $p_s(r)=p_s(1/r)$ with $r=1$ a fixed point. Precisely for $r=1$ we have minimal purity 
$p_s(1)=(\um(4s-1))!/(\sqrt{\pi} (2s)!)$, to be compared with the purity $1/(2s+1)$ of a maximally entangled state. 
\begin{figure}
\includegraphics[width=8cm]{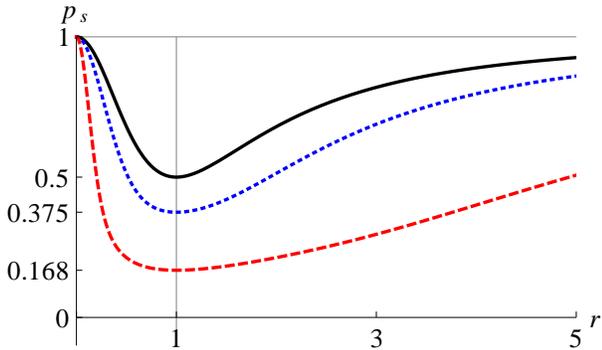}
\caption{ \label{puritysu2} (Color online) Purity $p_s(r)$  of the reduced density matrix $\varrho_b=\tr_a(\varrho)$
to layer $b$, for the $U(2)$ CS density matrix $\varrho=|z\rangle\langle z|$, as a function of the coherent
state parameter $r=|z|$ for three values of the pseudospin: $s=1/2$ (black), $s=1$ (dotted blue) and $s=11/2$ (dashed red).}
\end{figure}
In Figure 
\ref{puritysu2} we represent $p_s$ as a function of $r$ for several pseudospins. We see that $|z\rangle$ at $r=1$ is maximally entangled 
for $s=1/2$ since $p_{1/2}(1)=1/2$ (purity reaches its minimum). For higher pseudospin $s$ values we have the asymptotic behavior 
$p_s(1)=1/\sqrt{2\pi s}+ O(s^{-3/2})$ which says that the corresponding CS is never maximally entangled. 
The horizontal grid line of Figure 
\ref{puritysu2} indicates the pure-state purity, which is attained at $r=0$ [all particles in layer $b$, with CS $|z_0\rangle=|k=-s\ra$]  
and when $r\to\infty$  [all particles in layer $a$, with CS $|z_\infty\rangle=|k=s\ra$].

\begin{figure}
\includegraphics[width=8cm]{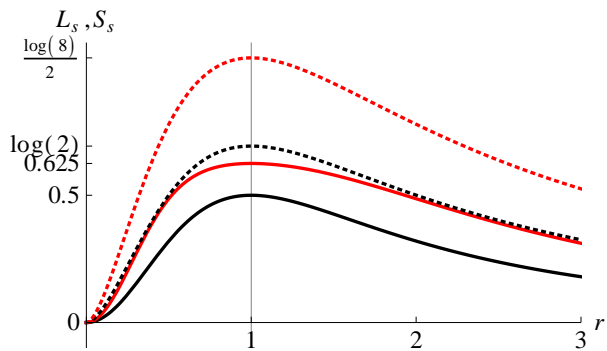}
\caption{ \label{Vonsu2} (Color online) Linear $L_s$ (solid line) and Von Neumann $S_s$ (dotted line) entanglement entropies 
of the reduced density matrix $\varrho_b=\tr_a(\varrho)$
to layer $b$, for the pseudospin-$s$ CS density matrix $\varrho=|z\rangle\langle z|$, as a function of the coherent
state parameter $r=|z|$ for two values of the pseudospin: $s=1/2$ (black) and $s=1$ (red).}
\end{figure}

For those readers more familiar with Von Neumann entropy $S_s(r)=-\tr(\varrho_b\log\varrho_b)=-\sum_{n=0}^{2s}\gamma_n(r)\log \gamma_n(r)$ we 
plot it in Figure \ref{Vonsu2} together with the linear entropy $L_s(r)=1-p_s(r)$, which turns out to be a lower approximation of $S_s$ (they are 
almost equal when the state is almost pure). We see that Von Neumann entropy is also maximum at $r=1$ and attains its maximum value 
$\log(2s+1)$ (completely mixed state) only for $s=1/2$, for which $S_{1/2}(1)=1/2$ [in general $S_s(1)\leq \log(2s+1)$]. 

In what follows, we shall not make the assumption that Zeeman energy is very large and we shall study how spin affects 
interlayer coherence and entanglement in a $U(4)$ symmetry setting (an intermediate step studying entanglement in $SU(2)\times SU(2)$ 
mixed bipartite quantum states has been considered in \cite{Schliemannsu22}). 
In Sec. \ref{entangsec} we shall also consider the interlayer entanglement of the (mixed, non-pure) 
equilibrium state of a BLQH spinning system at finite temperature. 

\section{U(4) operators and Hilbert space}\label{sec2}

Bilayer quantum Hall (BLQH) systems underlie an isospin $U(4)$ symmetry. In order to emphasize
the spin $SU(2)$ symmetry in the, let us say,  bottom $b$ (pseudospin down) and top $a$ or (pseudospin up)
layers, it is customary to denote the $U(4)$ generators in the four-dimensional fundamental representation by 
the sixteen $4\times 4$ matrices $\tau_{\mu\nu}\equiv\sigma_\mu\otimes\sigma_\nu, \, \mu,\nu=0,1,2,3$. The spin-frozen 
annihilation operators $a$ and $b$ will be replaced by their $2\times 2$ matrix counterparts $a\to\mathbf a$ and $b\to \mathbf b$, so that 
the two component ``field'' $\mathcal Z$ is now arranged as as a compound $\mathcal Z=(\mathcal Z_1,\mathcal Z_2)$ of two fermions as 
\be
\mathcal Z=\begin{pmatrix}
            \mathbf a\\ \mathbf b
           \end{pmatrix}=\begin{pmatrix}
\begin{matrix} a_1^\downarrow & a_2^\downarrow \\ a_1^\uparrow & a_2^\uparrow
\end{matrix}
\\ \begin{matrix} b_1^\uparrow & b_2^\uparrow\\ b_1^\downarrow & b_2^\downarrow
\end{matrix} \end{pmatrix}=
\begin{pmatrix}
\begin{matrix} a_0 & a_1\\ a_2 & a_3
\end{matrix}
\\ \begin{matrix} b_0 & b_1\\ b_2 & b_3
\end{matrix} \end{pmatrix}.\label{calzeta}\ee
The operator $(a_1^\downarrow)^\dag=a_0^\dag$ [resp. $(b_2^\uparrow)^\dag=b_1^\dag$] creates a flux quanta attached to the first [resp. second] electron
with spin down [resp. up] at layer $a$ [resp. $b$], and so on. We shall use the more compact notation $a_\mu, b_\mu, \mu=0,1,2,3$, and just remember that even
and odd quanta are attached to the first and second electrons, respectively. Note that the modes $\mu=\{0,1\}$ (resp. $\mu=\{2,3\}$) are related to spin up (resp. down)
in layer $b$ and viceversa in layer $a$; this is due to an inherent conjugated response of spin in each layer under $U(4)$ rotations 
[see later in paragraph before eq. \eqref{basislayer}]. The sixteen $U(4)$ isospin density operators [the spinning counterpart of \eqref{spinfrozen}] are then 
written as 
\be
\mathcal{T}_{\mu\nu}=\tr({\mathcal Z}^\dag\tau_{\mu\nu} \mathcal Z),\label{bosrepre} \ee
which constitute an oscillator representation of the $4\times 4$ matrix generators $\tau_{\mu\nu}$, in terms of eight bosonic modes. 
In a previous article \cite{GrassCSBLQH} we have obtained the matrix elements of $\mathcal{T}_{\mu\nu}$ in a Fock state basis.

In the previous Section, we fixed the total number of flux quanta $\mathcal Z^\dag\mathcal Z=a^\dag a+b^\dag b$ to $2s$. 
The $U(4)$ analogue of this constraint adopts the compact form
\[\mathcal Z^\dag \mathcal Z|{}{}_{q_a,q_b}^{j,m}\rangle=(\mathbf a^\dag \mathbf a+\mathbf b^\dag \mathbf b)|{}{}_{q_a,q_b}^{j,m}\rangle=\lambda
I_{2}|{}{}_{q_a,q_b}^{j,m}\rangle,\]
valid for any physical state $|{}{}_{q_a,q_b}^{j,m}\rangle$, 
where by $I_{2}$ we denote the $2\times 2$ identity operator and $\lambda$ is the number of flux quanta attached to each electron. In particular,
the linear Casimir operator $\mathcal{T}_{00}=\tr(\mathcal Z^\dag \mathcal Z)=\sum_{\mu=0}^3 a^{\dag}_\mu a_\mu+b^{\dag}_\mu b_\mu$, providing the total number of quanta,
is fixed to $2\lambda$, $\lambda\in\mathbb N$. We also identify the interlayer imbalance operator now as 
$\mathcal{T}_{30}=\sum_{\mu=0}^3 a^{\dag}_\mu a_\mu-b^{\dag}_\mu b_\mu$, which measures the excess of quanta between
layers $a$ and $b$. In the BLQH literature (see e.g.
\cite{EzawaBook}) it is customary to denote the total spin $\mathcal{S}_j=\mathcal{T}_{0j}/2$ and
pseudospin $\mathcal{P}_j=\mathcal{T}_{j0}/2$, together with the remaining 9 isospin  $\mathcal{R}_{kj}=\mathcal{T}_{jk}/2$ operators.

It is clear that \eqref{bosrepre} defines a unitary bosonic representation of the $U(4)$ matrix generators $\tau_{\mu\nu}$ in the Fock
space expanded by orthonormal basis states [the $U(4)$ analogue of \eqref{bosrepreu2}]
\be
\left|\begin{matrix} n_a^0 & n_a^1\\ n_a^2 & n_a^3
\end{matrix}\right>_a\!\!\otimes \left|\begin{matrix} n_b^0 & n_b^1\\ n_b^2 & n_b^3
\end{matrix}\right>_b\!\!=\prod_{\mu=0}^3
\frac{(a^\dag_\mu)^{n_a^\mu}(b^\dag_\mu)^{n_b^\mu}}{\sqrt{n_a^\mu!n_b^\mu!}}|0\ra,\label{grassmannbasis2}
\ee
where $|0\ra$ denotes the Fock vacuum and $n_\ell^\mu$ the occupancy numbers of layers $\ell=a,b$ and modes $\mu=0,1,2,3$.
The fact that the total number of quanta is constrained to  $\sum_{\mu=0}^3 n_a^\mu+n_b^\mu=2\lambda$ indicates that the representation
\eqref{bosrepre} is reducible in Fock space. 
In Ref. \cite{GrassCSBLQH} we have obtained the  carrier Hilbert space
${\cal H}_\lambda(\mathbb G_2)$ of a
\be d_\lambda=\frac{1}{12}(\lambda+1)(\lambda+2)^2(\lambda+3)\label{dimension}\ee
dimensional irreducible representation of $U(4)$ spanned by the
set of orthonormal basis vectors
\be
\left\{|{}{}_{q_a,q_b}^{j,m}\ra, \begin{matrix}
  2j, m\in\mathbb N,\\  q_a,q_b=-j,\dots,j \end{matrix}\right\}_{2j+m\leq\lambda}\label{basisvec}
\ee
in terms of Fock states \eqref{grassmannbasis2} (see Appendix \ref{basissubsec} for a brief).  These basis vectors fulfill a resolution of the identity
\be 1=\sum^{\lambda}_{m=0}\sum_{j=0;\um}^{(\lambda-m)/2}\sum^{j}_{q_a,q_b=-j}
|{}{}_{q_a,q_b}^{j,m}\ra \la{}{}_{q_a,q_b}^{j,m}|,\nn
\ee
where $\sum_{j=0;\um}$  means sum on $j=0,\um,1,\frac{3}{2},\dots$  
These are the $U(4)$ isospin-$\lambda$ analogue of the pseudospin-$s$  orthonormal 
basis vectors \eqref{basisinfocksu2}, with the role of the 
pseudospin $s$ played now by $\lambda$ (we are omitting the labels $s$ and $\lambda$ from the basis vectors $|k\rangle$ and $|{}{}_{q_a,q_b}^{j,m}\ra$, respectively, 
for the sake of brevity). Piercing the two electrons with $\lambda$ magnetic flux lines affects the total angular momentum $j$ of the system, 
which can reach the values $j=0,\um,1,\frac{3}{2},\dots,\frac{\lambda}{2}$. The meaning of the quantum (natural) number $m$ in 
$|{}{}_{q_a,q_b}^{j,m}\ra$ is related to the total number of flux quanta in layer $a$ by
\be
n_a^0+n_a^1+n_a^2+n_a^3=2(j+m)
\ee
and also to the  interlayer
imbalance $(2j+2m-\lambda)$ (the eigenvalue of the pseudospin third component $\mathcal{P}_3=\mathcal{T}_{30}/2$), measuring half the
excess of flux quanta in layer $a$ w.r.t. layer $b$. 
Note that $j$ and $m$ are always bounded by $2j+m\leq \lambda$, as stated in eq. \eqref{basisvec}, thus leading to a 
finite-dimensional representation of $U(4)$. 
The remainder quantum numbers $q_a$ and $q_b$ represent the angular momentum third components of layers $a$ and $b$, respectively. 
Their relation with flux quanta turns out to be:
\bea
n_a^0+n_a^1-n_a^2-n_a^3&=&-2q_a\,,\nn\\
n_b^0+n_b^1-n_b^2-n_b^3&=&2q_b\,,
\eea
which says that the ``spin third component quantum number''  $q_b$ measures the imbalance between $\mu=\{0,1\}$ (spin up) and
$\mu=\{2,3\}$ (spin down) type flux quanta  inside layer $b$, and viceversa for $q_a$ inside layer $a$ 
[remember the assignment in \eqref{calzeta}].

The explicit construction of the basis states \eqref{basisvec} in Fock state for general $\lambda$ entails a certain level of mathematical sophistication and has been 
worked out in a previous Ref. \cite{GrassCSBLQH}. In this Section we shall only discus the simpler case $\lambda=1$. This should be enough 
in a first reading. For the case of arbitrary $\lambda$, we have included a brief in Appendix \ref{basissubsec}, in order to make the 
presentation simpler and self-contained.

As the simplest example, let us provide the explicit expression of the  basis states $|{}{}_{q_a,q_b}^{j,m}\ra$ in terms 
of Fock states \eqref{grassmannbasis2} for
two flux quanta ($\lambda=1$ line of flux):
\bea
|{}{}_{0,0}^{0,0}\ra&=&\frac{1}{\sqrt{2}}\left(
\left|\begin{matrix} 0 & 0\\ 0 & 0
\end{matrix}\right>_a\!\!\!\otimes \left|\begin{matrix} 1 & 0\\ 0 & 1
\end{matrix}\right>_b-
\left|\begin{matrix} 0 & 0\\ 0 & 0
\end{matrix}\right>_a\!\!\!\otimes  \left|\begin{matrix} 0 & 1\\ 1 & 0
\end{matrix}\right>_b
\right),\nn\\
|{}{}_{\um,\um}^{\um,0}\ra&=&\frac{1}{\sqrt{2}}\left(
\left|\begin{matrix} 0 & 0\\ 0 & 1
\end{matrix}\right>_a\!\!\!\otimes \left|\begin{matrix} 1 & 0\\ 0 & 0
\end{matrix}\right>_b-
\left|\begin{matrix} 0 & 0\\ 1 & 0
\end{matrix}\right>_a\!\!\!\otimes  \left|\begin{matrix} 0 & 1\\ 0 & 0
\end{matrix}\right>_b
\right),\nn\\
|{}{}_{\frac{-1}{2},\frac{-1}{2}}^{\;\;\um,\;\;0}\ra&=&\frac{1}{\sqrt{2}}\left(
\left|\begin{matrix} 1 & 0\\ 0 & 0
\end{matrix}\right>_a\!\!\!\otimes \left|\begin{matrix} 0 & 0\\ 0 & 1
\end{matrix}\right>_b-
\left|\begin{matrix} 0 & 1\\ 0 & 0
\end{matrix}\right>_a\!\!\!\otimes  \left|\begin{matrix} 0 & 0\\ 1 & 0
\end{matrix}\right>_b
\right),\nn\\
|{}{}_{\frac{-1}{2},\um}^{\;\um,\,0}\ra&=&\frac{1}{\sqrt{2}}\left(
\left|\begin{matrix} 1 & 0\\ 0 & 0
\end{matrix}\right>_a\!\!\!\otimes \left|\begin{matrix} 0 & 1\\ 0 & 0
\end{matrix}\right>_b-
\left|\begin{matrix} 0 & 1\\ 0 & 0
\end{matrix}\right>_a\!\!\!\otimes  \left|\begin{matrix} 1 & 0\\ 0 & 0
\end{matrix}\right>_b
\right),\nn\\
|{}{}_{\um,\frac{-1}{2}}^{\,\um,\;0}\ra&=&\frac{1}{\sqrt{2}}\left(\left|\begin{matrix} 0 & 0\\ 0 & 1
\end{matrix}\right>_a\!\!\!\otimes  \left|\begin{matrix} 0 & 0\\ 1 & 0
\end{matrix}\right>_b-
\left|\begin{matrix} 0 & 0\\ 1 & 0
\end{matrix}\right>_a\!\!\!\otimes \left|\begin{matrix} 0 & 0\\ 0 & 1
\end{matrix}\right>_b
\right),\nn\\
|{}{}_{0,0}^{0,1}\ra&=&\frac{1}{\sqrt{2}}\left(\left|\begin{matrix} 1 & 0\\ 0 & 1
\end{matrix}\right>_a\!\!\!\otimes  \left|\begin{matrix} 0 & 0\\ 0 & 0
\end{matrix}\right>_b-
\left|\begin{matrix} 0 & 1\\ 1 & 0
\end{matrix}\right>_a\!\!\!\otimes \left|\begin{matrix} 0 & 0\\ 0 & 0
\end{matrix}\right>_b
\right).\nn\\ \label{basisl1}
\eea
This irreducible representation arises in the Clebsch-Gordan decomposition of a tensor product of $2$ four-dimensional (elementary) 
representations of $U(4)$ 
\be  \begin{Young}  \cr \end{Young}  \otimes \begin{Young}  \cr \end{Young}  =
  \begin{Young} & \cr \end{Young}\oplus \begin{Young} \cr \cr \end{Young}\,\Rightarrow\;\;
 4\times 4=10+6\nn
\ee
and corresponds to the totally antisymmetric case with dimension $d_1=6$, in accordance with \eqref{dimension}. 
It agrees with the fact that two electrons in one Landau site must form an
antisymmetric state due to Pauli exclusion principle. The $d_1=6$-dimensional irrep of $SU(4)$ is usually divided into two sectors
(see e.g. \cite{EzawaBook}): the spin sector with spin-triplet pseudospin-singlet states
\be |\mathfrak{S}_\uparrow\rangle=|{}{}_{\frac{-1}{2},\um}^{\,\um,\;0}\ra,\,
|\mathfrak{S}_0\rangle=\frac{1}{\sqrt{2}}(|{}{}_{\um,\um}^{\,\um, 0}\ra+|{}{}_{\frac{-1}{2},\frac{-1}{2}}^{\;\um,\;0}\ra),
\,|\mathfrak{S}_\downarrow\rangle=|{}{}_{\um,\frac{-1}{2}}^{\,\um,\;0}\ra\label{st} \ee
and the ppin sector with pseudospin-triplet  spin-singlet  states
\be|\mathfrak{P}_\uparrow\rangle=|{}{}_{0,0}^{0, 1}\ra,\,
|\mathfrak{P}_0\rangle=\frac{1}{\sqrt{2}}(|{}{}_{\um,\um}^{\um,\;0}\ra-|{}{}_{\frac{-1}{2},\frac{-1}{2}}^{\;\um,\;0}\ra),
\,|\mathfrak{P}_\downarrow\rangle=|{}{}_{0,0}^{0, 0}\ra.\label{pt} \ee

For arbitrary $\lambda$, the Young tableau of the corresponding $d_\lambda$-dimensional
representation is made of two rows of $\lambda$ boxes each. We can think of the following ``composite bi-fermion'' picture (following 
Jain's image \cite{Jainbook}) to
physically explain the dimension \eqref{dimension} of the Hilbert space ${\cal H}_\lambda(\mathbb
G_2)$. We have two electrons attached to $\lambda$ flux quanta each.
The first electron can occupy any of the four isospin states $|b\!\uparrow\rangle, |b\!\downarrow\rangle,
|a\!\uparrow\rangle$ and $|a\!\downarrow\rangle$ at one Landau site of the lowest Landau level. Therefore, there are $\tbinom{4+\lambda-1}{\lambda}$ ways of
distributing $\lambda$ quanta among these four states. Due to the Pauli exclusion principle, there are only three states
left for the second electron and $\tbinom{3+\lambda-1}{\lambda}$ ways of
distributing $\lambda$ quanta among these three states. However, some of the previous configurations must be identified
since both electrons are indistinguishable and $\lambda$ pairs of quanta adopt $\tbinom{2+\lambda-1}{\lambda}$
equivalent configurations. In total, there are
\begin{equation}
 \frac{\binom{\lambda+3}{\lambda}\binom{\lambda+2}{\lambda}}{\binom{\lambda+1}{\lambda}}=\frac{1}{12}(\lambda+3)(\lambda+2)^2(\lambda+1)\nn
\end{equation}
ways to distribute $2\lambda$ flux quanta among two identical electrons in four states, which turns out to coincide with
the dimension $d_\lambda$ in \eqref{dimension} of the Hilbert space  ${\cal H}_\lambda(\mathbb
G_2)$. Other possible picture, compatible with some usual ``flux line'' representations (see e.g. \cite{EzawaBook}), is the following.
We have $\lambda$ magnetic flux lines piercing two electrons which can occupy four possible states. There are
$\tbinom{\lambda+3}{\lambda}$ and $\tbinom{\lambda+2}{\lambda}$ ways of piercing the first and second electron, respectively.
Indistinguishability identifies $\tbinom{\lambda+1}{\lambda}$ of these possible configurations, 
rendering again $d_\lambda$ ways to pierce the two electrons with $\lambda$ flux lines.

Concerning the quantum statistics of our states for a given number $\lambda$ of flux lines, 
one can see that whereas the orthonormal basis functions \eqref{basisvec} [see also \eqref{basisinfock2}] are antisymmetric (fermionic character) under the 
interchange of the two electrons for $\lambda$ odd, they are symmetric (bosonic character) for $\lambda$ even. Indeed, under the 
interchange of columns in [interchange of electrons $(\mathcal{Z}_1,\mathcal{Z}_2)\to (\mathcal{Z}_2,\mathcal{Z}_1)$ in \eqref{calzeta}]  
\be \mathbf{a}=\begin{pmatrix} a_0 & a_1\\ a_2 & a_3 \end{pmatrix}\to \; 
\widetilde{\mathbf{a}}=\begin{pmatrix} a_1 & a_0\\ a_3 & a_2 \end{pmatrix},\label{atilde}\ee
[likewise for $\mathbf{b}\to\widetilde{\mathbf{b}}$], 
the basis states transform according to  (see Appendix \ref{basissubsec} for more details) 
\be\widetilde{|{}{}_{q_a,q_b}^{j,m}\ra}=(-1)^\lambda\, |{}{}_{q_a,q_b}^{j,m}\ra.\label{statistics}\ee
Indeed, it can be straightforwardly verified for $\lambda=1$ by swapping columns of vectors of $|\cdot\ra_a$ and $|\cdot\ra_b$ 
in \eqref{basisl1} and, in general, by exchanging  columns in the Fock states \eqref{grassmannbasis2}. 
There is another inherent symmetry under the interchange of layers $a$ and $b$, although it does not entail any kind of quantum statistics because 
layers $a$ and $b$ are not  indistinguishable.  
In fact, from the general definition of the basis states $|{}{}_{q_a,q_b}^{j,m}\ra$ in eq. \eqref{basisinfock2}, it is easy to see that, under the 
interchange of layers $\mathbf{a}\leftrightarrow \mathbf{b}$, we have the following ``population-inversion'' property 
\be
|{}{}_{q_a,q_b}^{j,m}\ra \leftrightarrow (-1)^{q_a-q_b}|{}{}_{-q_b,-q_a}^{j,\lambda-2j-m}\ra,
\ee
which says that the population of layer $a$, $p_a=2j+2m$, becomes $2j+2(\lambda-2j-m)=2\lambda-p_a=p_b$, the population of layer $b$. 
Indeed, one can check this property for the easier $\lambda=1$ case directly in \eqref{basisl1}. 
We shall see that interlayer entanglement depends quantitatively on $\lambda$ (see e.g. Figure \ref{purity}), but the qualitative behavior turns out to be similar 
for $\lambda$ even (bosonic) and odd (fermionic). This is because we are studying layer-layer entanglement but not fermion-fermion or boson-boson 
entanglement, for which a dependence on the parity of $\lambda$ is expected. Moreover, when discussing layer-layer entanglement we do not have to worry 
about filtering the intrinsic correlations of identical particles due to Pauli exclusion principle (see 
e.g. \cite{entangidpart}), since layers are not indistinguishable.

In Ref. \cite{GrassCSBLQH} (see the Appendix \ref{subsecCS} for a brief) we have also introduced a set of 
(quasi-classical) CS with interesting mathematical and also physical properties that will be analyzed here and in future 
publications. These CS constitute a kind of matrix generalization of the pseudospin-$s$ CS \eqref{su2cs}. They are 
also Bose-Einstein-like condensates [see eq. \eqref{u4csfock}] but they are now labeled 
by a $2\times 2$ complex matrix $Z=z^\mu\sigma_\mu$ (sum on $\mu=0,1,2,3$), with four
complex entries  $z^\mu=\tr(Z\sigma_\mu)/2$ (they are points on the eight-dimensional Grassmannian $\mathbb{G}_2$).  
These CS  can be expanded in terms of the orthonormal basis vectors \eqref{basisvec}, and the general 
formula is given in \eqref{u4cs}. In this Section we just shall  write the expression of $|Z\ra$ for the simplest $\lambda=1$ case 
in terms of the basis states \eqref{basisl1}:
\bea
|Z\ra&=&\left[|{}{}_{0,0}^{0,0}\ra+
\det(Z)|{}{}_{0,0}^{0,1}\ra+
(z^0-z^3)|{}{}_{\frac{-1}{2},\frac{-1}{2}}^{\;\;\um,\;\;0}\ra+\right.\nn\\ 
&& \left. (z^1+iz^2)|{}{}_{\frac{-1}{2},\um}^{\;\um,\,0}\ra+
(z^1-iz^2)|{}{}_{\um,\frac{-1}{2}}^{\,\um,\;0}\ra+\right.\nn\\ && 
\left. (z^0+z^3)|{}{}_{\um,\um}^{\um,0}\ra\right]/
\det(\sigma_0+Z^\dag Z)^\um,\label{cs1}
\eea
where the denominator is a normalizing factor. In therms of the spin-triplet \eqref{st} and 
pseudospin triplet \eqref{pt} states, we  equivalently have 
\bea
|Z\ra&=&\left[ |\mathfrak{P}_\downarrow\rangle+\det(Z)|\mathfrak{P}_\uparrow\rangle+\sqrt{2} 
z^3|\mathfrak{P}_0\rangle+
\right.\nn\\ 
&& \left. z^1(|\mathfrak{S}_\uparrow\rangle+|\mathfrak{S}_\downarrow\rangle) 
+i z^2(|\mathfrak{S}_\uparrow\rangle-|\mathfrak{S}_\downarrow\rangle)+\right.\nn\\
&&\left. \sqrt{2}z^0|\mathfrak{S}_0\rangle\right]/
\det(\sigma_0+Z^\dag Z)^\um.\label{cs11}
\eea
Therefore, the CS $|Z\ra$ depends on four arbitrary complex parameters $z^\mu, \mu=0,1,2,3$ [and not only one $z$ like the spin-frozen 
case \eqref{su2cs}], which means that we have extra isospin operators to create interlayer and spin coherence.  
A particular experimental way to generate these CS is through the natural tunneling interaction
arising when both layers are placed close enough and electrons hop between them [see formula \eqref{u4cs2},  
which provides an expression of the CS $|Z\ra$  as an exponential of interlayer ladder operators 
$\mathcal{T}_{\pm\mu}=(\mathcal{T}_{1\mu}\pm i\mathcal{T}_{2\mu})/2$].

In the next two sections, we shall study 
some physical quantities that only depend on two (out of the eight $z^\mu$) parameters related to the 
determinant $\det(ZZ^\dag)$ and trace $\tr(ZZ^\dag)$, 
due to an intrinsic rotational invariance.

\section{Interlayer coherence  and imbalance fluctuations}\label{secimb}

\begin{figure}
\includegraphics[width=8cm]{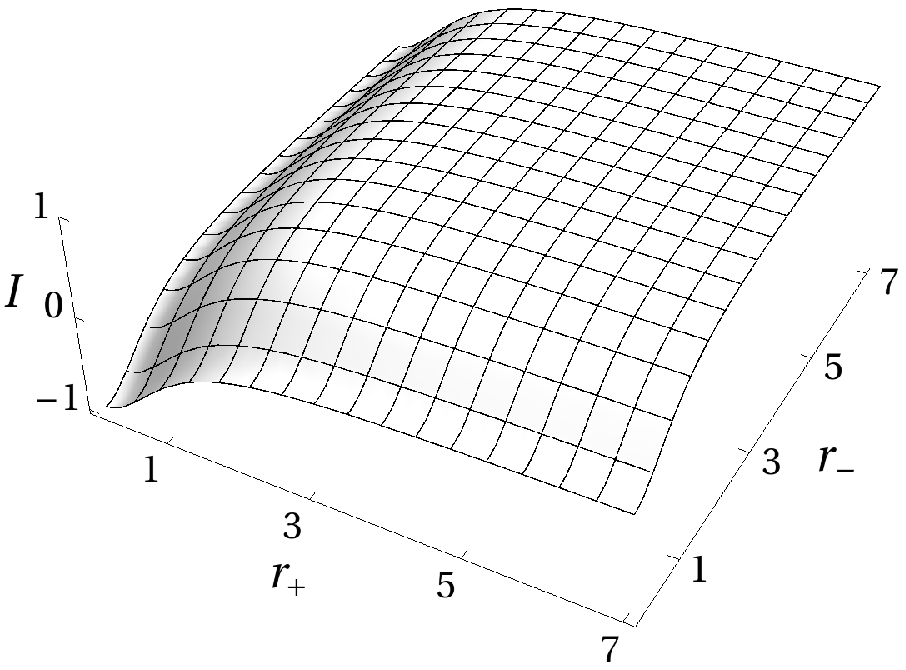}
\includegraphics[width=8cm]{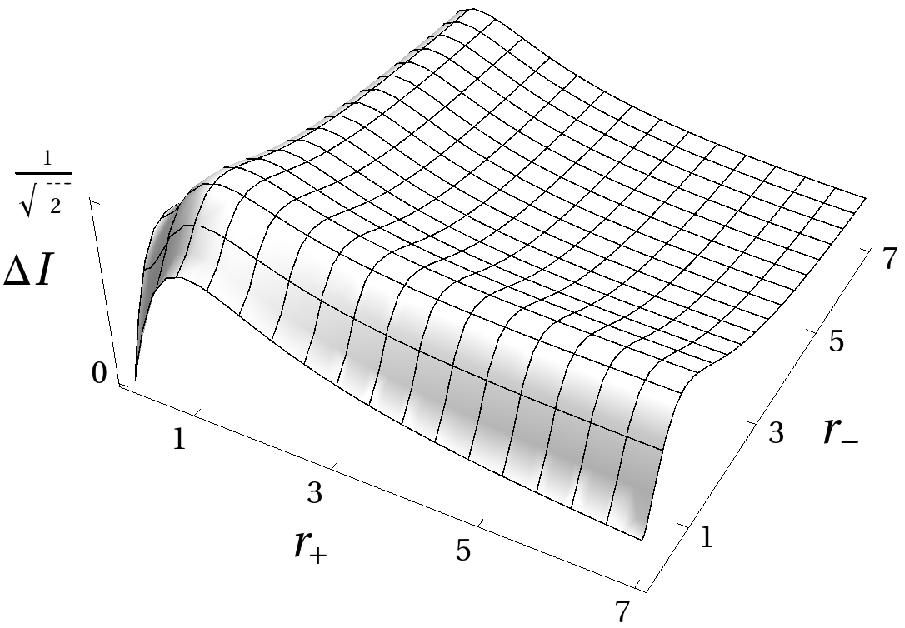}
\caption{ \label{imbalance3d} Imbalance $I$ and its standard deviation $\Delta I$ (per flux quanta) as a function of the
CS parameters $r_\pm$.}
\end{figure}

\begin{figure}
\includegraphics[width=8cm]{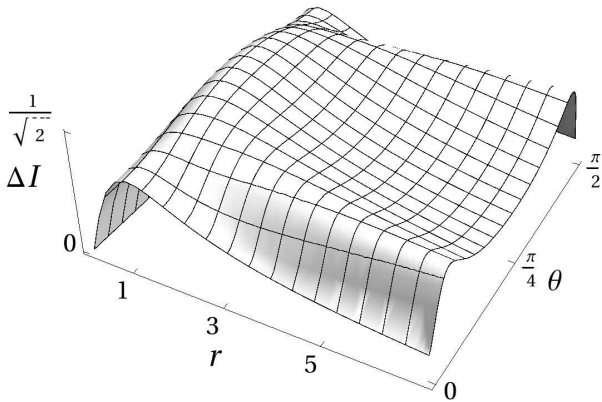}
\caption{ \label{imbalancedev3dpol} Standard deviation $\Delta I$ (per flux quanta) as a function of the CS 
parameters $r=\sqrt{r_+^2+r_-^2}$ and $\theta=\arctan(r_-/r_+)$.}
\end{figure}

\begin{figure}
\includegraphics[width=8cm]{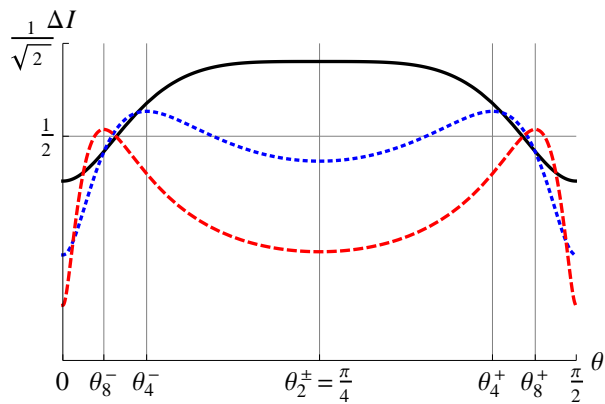}
\caption{ \label{imbalancedevtheta} (Color online) Standard deviation $\Delta I$ (per flux quanta) as a function of the polar angle
 $\theta=\arctan(r_-/r_+)$ for $r=2$ (solid line) $r=4$ (dotted blue) and $r=8$ (dashed red). The points $\theta^\pm_r$ denote
 local maxima of $\Delta I$ for each value of $r$}
\end{figure}

Like we did in Section \ref{imbsubsec} for the spin-frozen case, here we shall compute interlayer coherence and imbalance 
fluctuations but for the Grassmannian $\mathbb G_2$ CS $|Z\rangle$. Taking into account that
the basis state $|{}{}_{q_a,q_b}^{j,m}\ra$ is an eigenstate of $\mathcal{P}_3=\mathcal{T}_{30}/2$ with eigenvalue $(2j+2m-\lambda)$, 
we arrive at the following expression for its mean value in the CS $|Z\ra$ (we write the case of arbitrary $\lambda$):
\be
\langle \mathcal{P}_3\rangle=\langle Z|\mathcal{P}_3|Z\rangle=\lambda\frac{\det(Z^\dag Z)-1}{\det(\sigma_0+Z^\dag Z)}
\ee
Note that, since  $\det(\sigma_0+Z^\dag Z)=1+\tr(Z^\dag Z)+\det(Z^\dag Z)$, the mean value
$\la\mathcal{P}_3\rangle$ is only a function of the two $U(2)^2$-invariants: determinant $d=\det(Z^\dag Z)=\bar{z}^2z^2$ [with
$z^2\equiv {z}^\mu\eta_{\mu\nu}z^\nu$] and trace
$t=\tr(Z^\dag Z)=\bar{z}^\mu\delta_{\mu\nu}z^\nu$; we are using the Einstein summation convention with $\eta_{\mu\nu}=\mathrm{diag}(1,-1,-1,-1)$
the Minkowski metric and $\delta_{\mu\nu}=\mathrm{diag}(1,1,1,1)$ the Euclidean metric.  Instead of $d$ and $t$, we shall use other parametrization
adapted to the decomposition
\be
Z=V_a^\dag\begin{pmatrix} r_+e^{i\vartheta_+} & 0\\ 0 & r_-e^{i\vartheta_-} \end{pmatrix} V_b,\label{decompZ}\ee
where $V_a, V_b\in SU(2)$ are rotations, and $r_\pm\in[0,\infty)$ and
$\vartheta_\pm\in[0,2\pi)$ are polar coordinates. Taking into account that $d=r_+^2r_-^2$ and $t=r_+^2+r_-^2$,
the imbalance mean value ``per flux quanta'' is simply
\be
I(r_+,r_-)=\frac{\langle \mathcal{P}_3\rangle}{\lambda}=\frac{r_+^2r_-^2-1}{(1+r_+^2)(1+r_-^2)}.
\ee
In the same way, we can compute the imbalance variance ``per flux quanta'', which results in
\begin{eqnarray}
\Delta I^2(r_+,r_-)&=&\frac{\langle \mathcal{P}_3^2\rangle-\langle \mathcal{P}_3\rangle^2}{\lambda}\\
&=&\frac{r_+^2+r_-^2+4r_+^2r_-^2+r_+^4r_-^2+r_+^2r_-^4}{(1+r_+^2)^2(1+r_-^2)^2}.\nn
\end{eqnarray}
Note that $I$ and $\Delta I$ are independent of $\lambda$, since the mean value $\la\mathcal{P}_3\ra$
scales with $\lambda$ and its uncertainty
$\Delta\mathcal{P}_3$  scales with $\lambda^{1/2}$. Note also that $I$ and $\Delta I$ verify the following inversion
invariance
\be
I(r_+,r_-)=-I(\frac{1}{r_\mp},\frac{1}{r_\pm}),\; \Delta I(r_+,r_-)=\Delta I(\frac{1}{r_\mp},\frac{1}{r_\pm}).\label{imbinv}
\ee
In Figure \ref{imbalance3d} we
represent the imbalance $I$ and its standard deviation $\Delta I$ as a function of $r_\pm$.
We see that $I$ is an increasing function of $r_\pm$ and
takes its values in the interval $[-1,1]$. Balanced coherent configurations ($I=0$) occur on the hyperbola $r_+r_-=1$. The behavior of
$\Delta I$ is a bit more complex. The global maximum of  $\Delta I$ occurs at $r_+=r_-=1$, where the deviation attains the value $1/\sqrt{2}$. For
high values of $r_\pm$ the deviation $\Delta I$ tends to zero except for two particular trajectories. To better appreciate this fact, we use
polar coordinates $r_+=r\cos\theta$ and $r_-=r\sin\theta$, with $r\in[0,\infty)$ and $\theta\in[0,\pi/2]$. Figure \ref{imbalancedev3dpol}
offers a representation of $\Delta I$ as a function of $r$ and $\theta$ and Figure \ref{imbalancedevtheta} displays three sections (cuts)
($r=2, r=4$ and $r=8$) of $\Delta I$ as a function of $\theta$.
For $r\leq 2$, the $r=$constant cuts of the deviation $\Delta I$ have a single maximum at $\theta=\pi/4$. However, for $r>2$  the situation
changes and the cuts (for fixed $r$) of $\Delta I$ display two local maxima at two values of the polar angle $\theta^\pm_r$ given by
\be
\cos\theta^\pm_r=\sqrt{\frac{r^4\mp 2\sqrt{r^4-16}\mp r^2(\mp 4+\sqrt{r^4-16})}{2r^2(r^2+4)}}.\label{thetapm}
\ee
The expression  $\theta_\pm(r)=\theta^\pm_r$ gives
two singular trajectories in the $(r,\theta)$ plane for which fluctuations are always non-zero and tend to $\Delta I=1/2$ when
$r\to\infty$. Both local maxima are narrower and narrower (see Figure \ref{imbalancedevtheta}), with
$\theta^-_r\to 0$ and $\theta^+_r\to\pi/2$ as  as $r\to\infty$. We also have $\theta^\pm_2=\pi/4$.

Looking for a physical interpretation and implementation of interlayer coherence and imbalance fluctuations, let us
consider the case $Z=z^0\sigma_0+z^3\sigma_3$, for which $r_+=|z^0+z^3|$ and $r_-=|z^0-z^3|$, and the polar angle is then
$\theta=\arctan(|z^0-z^3|/|z^0+z^3|)$. According to the expression \eqref{u4cs2}, this CS can be generated by the operators
$\mathcal{T}_{\pm 0}$ and $\mathcal{T}_{\pm 3}$ or, equivalently, by $\mathcal{P}_1, \mathcal{P}_2$ and
 $\mathcal{R}_{31}, \mathcal{R}_{32}$ introduced
in Subsection \ref{sec2} to compare with the notation of standard textbooks like \cite{EzawaBook}.
The operators $\mathcal{P}_1$ and $\mathcal{P}_2$ produce the typical interlayer tunneling interaction present
in the spin-frozen Hamiltonian of the bilayer system. Here we have extra isospin operators  $\mathcal{R}_{31}$ and  $\mathcal{R}_{32}$
to play with to create interlayer coherence and imbalance. Actually, the peculiar situation described by the equation
\eqref{thetapm} takes place when tunneling interaction strengths  $z^0$ and $z^3$ of
$\mathcal{P}_1$ and  $\mathcal{R}_{31}$, respectively, verify $\tan\theta^\pm_r=|z^0-z^3|/|z^0+z^3|$ for each value of 
$r=\sqrt{2}\sqrt{|z^0|^2+|z^3|^2}$. For $r\leq 2$, maximum imbalance fluctuations occur when $r_+=r_-$, for example when $z^3=0$
(i.e. when the interaction $\mathcal{R}_{32}$ is switched off)
whereas for $r\gg 2$ maximum imbalance fluctuations require both tunneling interactions
to be slightly ``out of tune'', that is, when the corresponding tunneling interaction strengths $z^0$ and $z^3$ fulfill
$z^0\approx \pm z^3$. It would be worth to experimentally explore these situations.

Note that we recover the spin-frozen magnitudes as a particular case of
the general spinning case. In fact, this happens for the diagonal case   $r_+=r_-=r_0$, which corresponds to
$Z=z^0\sigma_0$ with $r_0=|z_0|$, where CS are just created by the tunneling interaction generated by $\mathcal{P}_{\pm}$,
discarding the extra $U(4)$ isospin generators $\mathcal{R}_{jk}$. The imbalance $I(r_0,r_0)$ and its standard deviation 
$\Delta I(r_0,r_0)$ for this case coincide with $\iota(r_0)$ and $\Delta\iota(r_0)$ in \eqref{imbsu2}. 

We again stress that the spinning case provides more degrees of freedom than the spin-frozen case to play with,
since we have extra isospin operators in $u(4)$ to create coherence. 
Actually, other isospin CS mean values $\la\mathcal{T}_{\mu\nu}\ra$, like 
the aforementioned interlayer phase difference $\theta(Z)=\arctan\la\mathcal{P}_2\ra/\la\mathcal{P}_1\ra$, 
will depend now on more that two CS parameters $z^\mu$. These 
cases deserve a separate study and will not be treated here.

We expect many more interesting physical phenomena at the previous critical points. Actually,
let us see that maximum interlayer entanglement also occurs at $r_0=1$ for a CS $|Z\ra$.

\section{Interlayer entanglement}\label{entangsec}

In the state space $\mathcal{H}_\lambda(\mathbb G_2)$, we shall consider the bipartite quantum system given by layers $a$ and $b$. 
Interlayer entanglement can provide feasible quantum computation. For example, in reference \cite{Scarola} it is 
theoretically shown that spontaneously interlayer-coherent BLQH droplets
should allow robust and fault-tolerant pseudospin quantum computation in semiconductor nanostructures. 
Here we shall show that BLQH coherent states at $\nu=2/\lambda$ are highly entangled, for high enough $\lambda$, and entanglement 
is robust (with low fluctuations) in a wide range of coherent state parameters.

\subsection{Interlayer entanglement of basis states}\label{entbasissec}
Let us firstly show that, contrary to the (direct product) basis states $|k\ra$ of the spin-frozen case in eq. \eqref{basisinfocksu2}, 
the orthonormal basis 
vectors $|{}{}_{q_a,q_b}^{j,m}\ra$ in \eqref{basisvec}   
are entangled for non-zero angular momentum, $j\not=0$. We shall explicitly work out the simplest case $\lambda=1$ and 
give the results for general $\lambda$. In the Appendix \ref{basissubsec} we show that the basis states 
$|{}{}_{q_a,q_b}^{j,m}\ra$ can be written as an expansion  
\be
|{}{}_{q_a,q_b}^{j,m}\ra=\sum_{q=-j}^{j}\frac{(-1)^{q_a-q}}{\sqrt{2j+1}}
|v_{-q,-q_a}^{j,m}\ra_a \otimes
|{v}_{q,q_b}^{j,\lambda-2j-m}\ra_b,\label{basisinfock1}
\ee
where $\{|v_{q,q'}^{j,m}\ra_a\}$ and
$\{|v_{q,q'}^{j,m}\ra_b\}$ are Schmidt basis for layers $a$ and $b$, respectively,
and $1/\sqrt{2j+1}$ are the Schmidt coefficients with
Schmidt number $2j+1$. For $\lambda=1$  the Schmidt (orthonormal) basis for layer 
$a$ (likewise for layer $b$) is simply: 
\bea
&&|v_{0,0}^{0,0}\ra_a=\left|\begin{matrix} 0 & 0\\ 0 & 0
\end{matrix}\right>_a,\, 
|v_{0,0}^{0,1}\ra_a=\frac{1}{\sqrt{2}}\left(\left|\begin{matrix} 1 & 0\\ 0 & 1
\end{matrix}\right>_a-\left|\begin{matrix} 0 & 1\\ 1 & 0
\end{matrix}\right>_a\right),\nn\\ 
&&|v_{\frac{-1}{2},\frac{-1}{2}}^{\;\;\um,\;\;0}\ra_a=\left|\begin{matrix} 0 & 0\\ 0 & 1
\end{matrix}\right>_a,\,
|v_{\frac{-1}{2},\um}^{\;\um,\,0}\ra_a=\left|\begin{matrix} 0 & 1\\ 0 & 0
\end{matrix}\right>_a,\nn\\
&& |v_{\um,\frac{-1}{2}}^{\,\um,\;0}\ra_a=\left|\begin{matrix} 0 & 0\\ 1 & 0
\end{matrix}\right>_a,\, 
|v_{\um,\um}^{\um,0}\ra_a=\left|\begin{matrix} 1 & 0\\ 0 & 0
\end{matrix}\right>_a.\label{Schmidt}
\eea
It can be easily checked that, plugging \eqref{Schmidt} into \eqref{basisinfock1}, one arrives to 
eq. \eqref{basisl1}.  For arbitrary $\lambda$, the Schmidt basis vectors for layer $a$ (idem for layer $b$) 
 are given in the Appendix \ref{basissubsec} by eq. \eqref{basislayer}, which fulfill the orthogonality relations \eqref{ortholayer}.

As a measure of interlayer entanglement, we shall compute the purity of the reduced density matrix (RDM) to one of the layers. 
More precisely, denoting by $\rho=|{}{}_{q_a,q_b}^{j,m}\ra\la{}{}_{q_a,q_b}^{j,m}|$ the density matrix of an arbitrary basis 
state and by $\rho_a=\tr_b(\rho)$ the reduced density matrix of layer $a$, it can be seen that 
the purity of $\rho_a$ is then $\tr(\rho_a^2)=1/(2j+1)$,
which is less than 1 if $j\not=0$. Indeed, the proof is apparent from the explicit expression of orthonormal basis vectors
$|{}{}_{q_a,q_b}^{j,m}\ra$  in \eqref{basisinfock1} and the fact that the Schmidt basis \eqref{Schmidt} is orthonormal for 
$\lambda=1$ [see \eqref{basislayer} and \eqref{ortholayer} for arbitrary $\lambda$]. Moreover,
tracing out the layer $b$ part, the reduced density matrix of layer $a$ is (we write the general $\lambda$ case) 
\begin{eqnarray}
 \rho_a&=&\tr_b(\rho)=\sum^{\lambda}_{m=0}\sum_{j=0;\um}^{(\lambda-m)/2}\sum^{j}_{q,q'=-j}
 {}_b\la v_{q,q'}^{j,m}|\rho|v_{q,q'}^{j,m}\ra_b\nn\\ &=&
 \frac{1}{2j+1}\sum_{q=-j}^{j}|v_{q,q_a}^{j,m}\ra_a\la v_{q,q_a}^{j,m}|\,.
\end{eqnarray}
Using again the orthonormality relations for Schmidt basis [see \eqref{ortholayer} for the general case], 
we finally arrive to the purity $\tr(\rho_a^2)=1/(2j+1)$. Therefore, for high angular momentum, $j\gg 1$, the basis state
$|{}{}_{q_a,q_b}^{j,m}\ra$ is highly entangled (almost zero purity) but not maximally entangled (with minimal purity $1/d_\lambda$), since $d_\lambda>2j+1$. In fact, 
we have that 
\be
\sum_{m=0}^\lambda \sum_{j=0;\um}^{(\lambda-m)/2}({2j+1})^2={d_\lambda}.\label{dimensionj}
\ee

\subsection{Interlayer entanglement of coherent states}

\begin{figure}
\includegraphics[width=8cm]{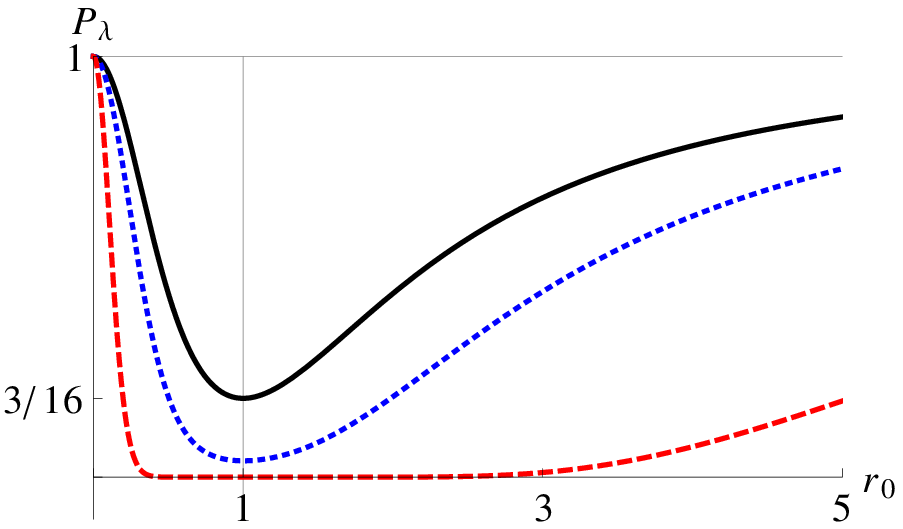}
\includegraphics[width=8cm]{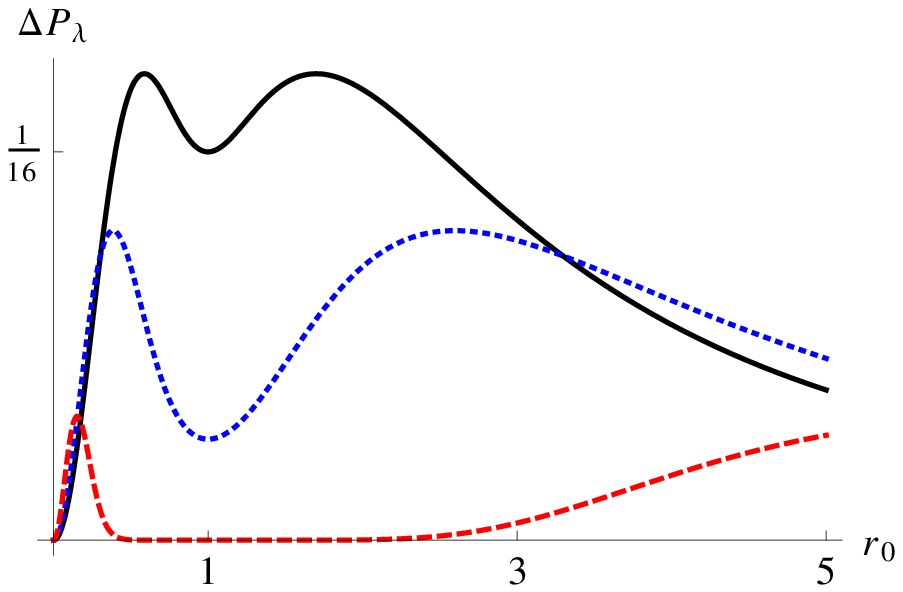}
\caption{ \label{purity} (Color online) Purity $P_\lambda$ and  standard deviation $\Delta P_\lambda$
of the reduced density matrix $\varrho_b=\tr_a(\varrho)$
to layer $b$, for the CS density matrix $\varrho=|Z\rangle\langle Z|$, as a function of the CS parameter
$r_0=r_+=r_-$ for three values of the total (half) number of flux quanta: $\lambda=1$ (solid), $\lambda=2$ (dotted blue) and $\lambda=11$ (dashed red).}
\end{figure}

Secondly, we shall study the interlayer entanglement of a CS 
$|Z\ra$. For example, for $\lambda=1$, and starting from \eqref{cs1} or \eqref{cs11}, it is 
relatively easy to see that the purity of the RDM $\varrho_b=\tr_a(\varrho)$ for $\varrho=|Z\ra\la Z|$ is
\be\tr(\varrho_b^2)_1=\frac{1+\um\tr(Z^\dag Z)^2-\det(Z^\dag Z)+\det(Z^\dag Z)^2}{{\det(\sigma_0+Z^\dag Z)^2}}.\label{purityl1}\ee
For arbitrary $\lambda$, calculations are more complicated and give the following expression for the RDM of layer $b$
\bea
\varrho_b&=&\sum_{m=0}^\lambda \sum_{j=0;\um}^{(\lambda-m)/2}\sum^{j}_{q,q',q_a,q_b=-j}\frac{1}{2j+1}\label{rdmcs}\\
&&\times \frac{\overline{\varphi_{-q,q_b}^{j,m}({Z})}\varphi_{-q,q_a}^{j,m}(Z)}{\det(\sigma_0+Z^\dag Z)^\lambda}
|v_{-q',q_a}^{j,\lambda-2j-m}\ra_b\la v_{-q',q_b}^{j,\lambda-2j-m}|,\nn
\eea
where $\varphi_{-q,q_b}^{j,m}({Z})$ are homogeneous polynomials of degree $2j+2m$ in $z^\mu$ given in  \eqref{basisfunc}. 
Using the orthonormality relations \eqref{ortholayer} for layer $b$, the purity can be finally written as
\be
\tr(\varrho_b^2)=\frac{\sum_{m=0}^\lambda \sum_{j=0;\um}^{(\lambda-m)/2}\sum^{j}_{q_{a,b}=-j}\frac{1}{2j+1}
\Phi_{q_a,q_b}^{j,m}({Z^\dag Z})}{\det(\sigma_0+Z^\dag Z)^{2\lambda}/\det(\sigma_0+(Z^\dag Z)^2)^{\lambda}}\label{purityb}
\ee
where we have defined the normalized probabilities
\be
\Phi_{q_a,q_b}^{j,m}({Z^\dag Z})= \frac{\overline{\varphi_{q_a,q_b}^{j,m}({Z^\dag Z})}\varphi_{q_a,q_b}^{j,m}(Z^\dag Z)}{\det(\sigma_0+(Z^\dag Z)^2)^\lambda}
\ee
which fulfill \[\sum_{m=0}^\lambda \sum_{j=0;\um}^{(\lambda-m)/2}\sum^{j}_{q_{a,b}=-j}\Phi_{q_a,q_b}^{j,m}({Z^\dag Z})=1.\]
This way of writing the purity \eqref{purityb} leads to an interesting physical interpretation of it. Note that
$\Phi_{q_a,q_b}^{j,m}({Z^\dag Z})$ is precisely the probability of finding the CS $|Z^\dag Z\ra$
in the basis state $|{}_{q_a,q_b}^{j,m}\ra$. Then the purity  can be written as the average value
\be \tr(\varrho_b^2)=\frac{ \langle Z^\dag Z|\frac{1}{2\mathcal{J}+1}| Z^\dag Z\rangle}
{\det(\sigma_0+Z^\dag Z)^{2\lambda}/\det(\sigma_0+(Z^\dag Z)^2)^{\lambda}},
\ee
with $j$ the eigenvalue of $\mathcal{J}$ with eigenvector $|{}_{q_a,q_b}^{j,m}\ra$.
From this point of view, we can also quantify purity fluctuations by defining the \emph{purity standard deviation} as
\be
\Delta \tr(\varrho_b^2)=\frac{\sqrt{\langle Z^\dag Z|\frac{1}{(2\mathcal{J}+1)^2}| Z^\dag Z\rangle-
\langle Z^\dag Z|\frac{1}{2\mathcal{J}+1}| Z^\dag Z\rangle^2}}
{\det(\sigma_0+Z^\dag Z)^{2\lambda}/\det(\sigma_0+(Z^\dag Z)^2)^{\lambda}}.
\ee
The physical meaning of purity variance is related to the robustness of entanglement, an important feature in feasible quantum computation. 
Low purity fluctuations are desirable when preparing entangled states low-sensitive to noise. We shall see that $|Z\rangle$ is almost maximally entangled 
in a wide range of CS parameters $Z$ with low purity variance (see later on Figure  \ref{purity}), specially for high values 
of the number of magnetic flux lines $\lambda$.

\begin{figure}
\includegraphics[width=8cm]{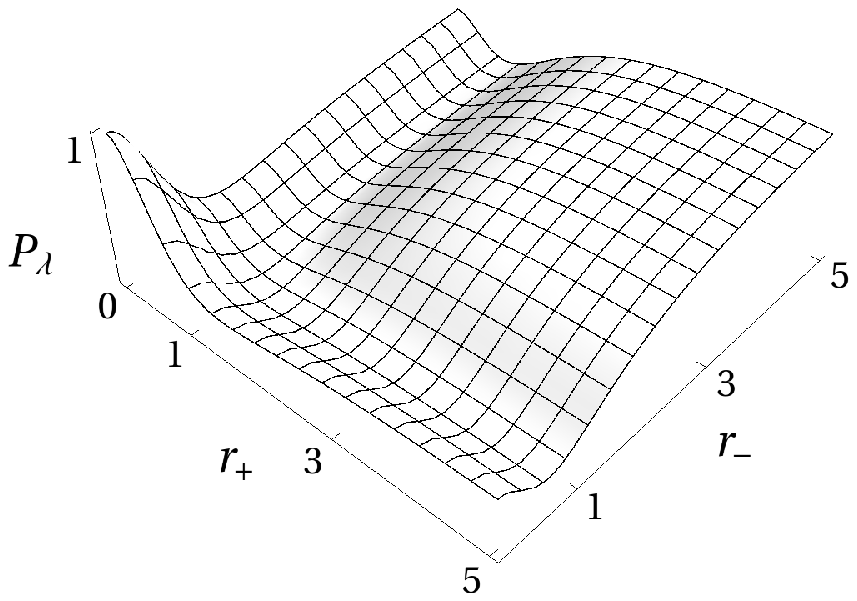}
\includegraphics[width=8cm]{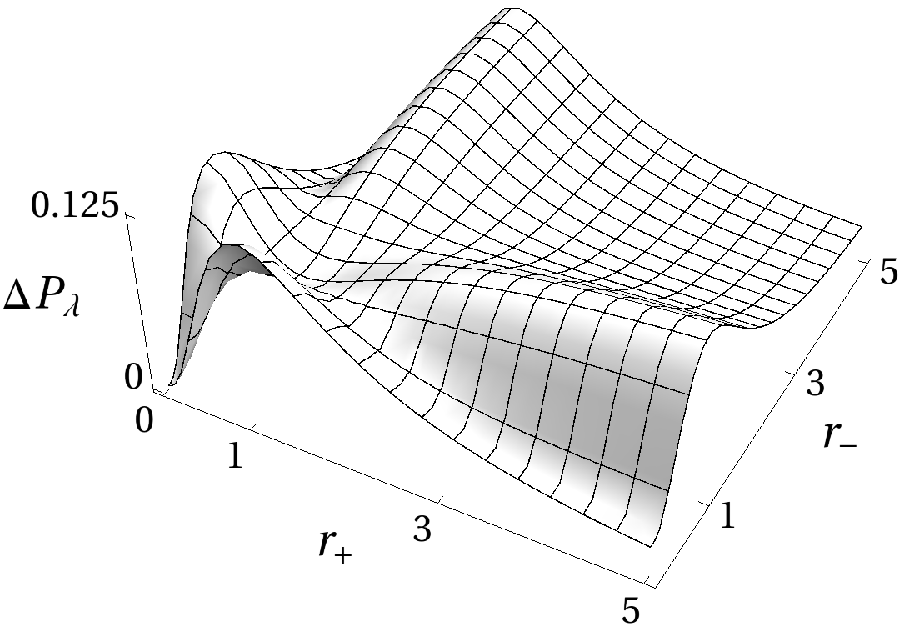}
\caption{ \label{purity13d} Purity $P_\lambda=\tr(\varrho_b^2)$, and  standard deviation $\Delta P_\lambda$,
of the reduced density matrix $\varrho_b=\tr_a(\varrho)$
to layer $b$, for the CS density matrix $\varrho=|Z\rangle\langle Z|$, as a function of the CS parameters
$r_\pm$ for $\lambda=1$.}
\end{figure}

\begin{figure}
\includegraphics[width=8cm]{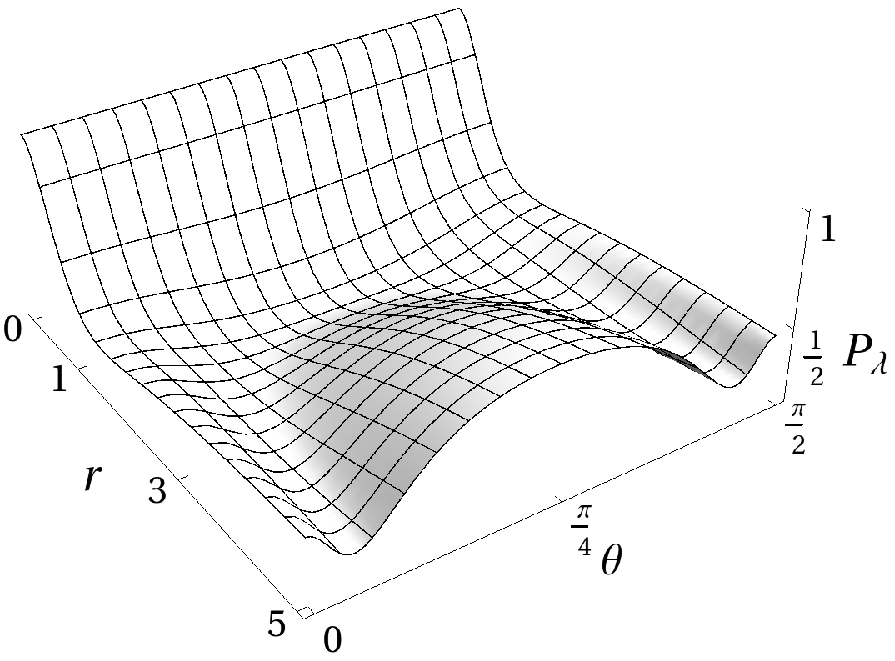}
\includegraphics[width=8cm]{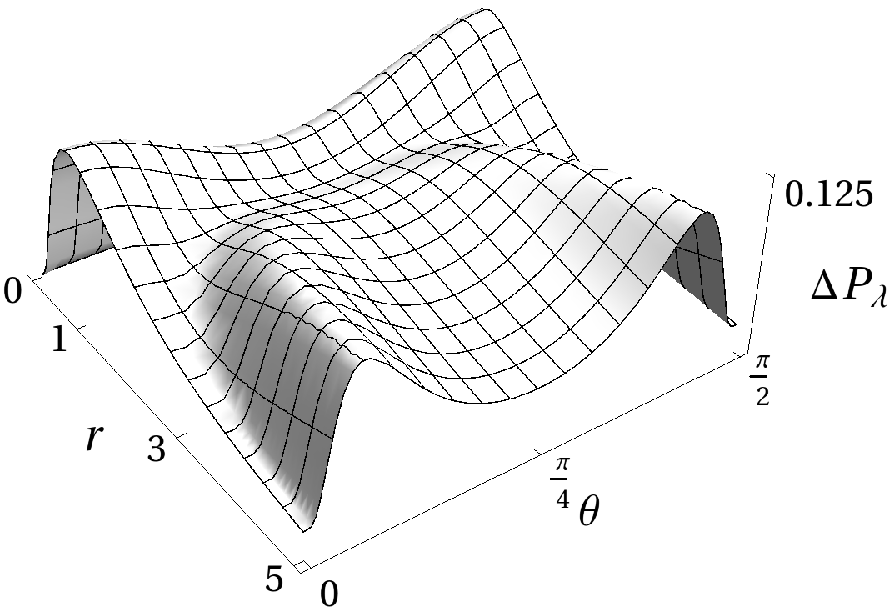}
\caption{ \label{purity13dpol} Purity $P_\lambda=\tr(\varrho_b^2)$, and  standard deviation $\Delta P_\lambda$,  of the reduced
density matrix $\varrho_b=\tr_a(\varrho)$
to layer $b$, for the CS density matrix $\varrho=|Z\rangle\langle Z|$, as a function of the CS parameters
$r=\sqrt{r_+^2+r_-^2},\theta=\arctan(r_-/r_+)$ for $\lambda=1$.}
\end{figure}

\begin{figure}
\includegraphics[width=8cm]{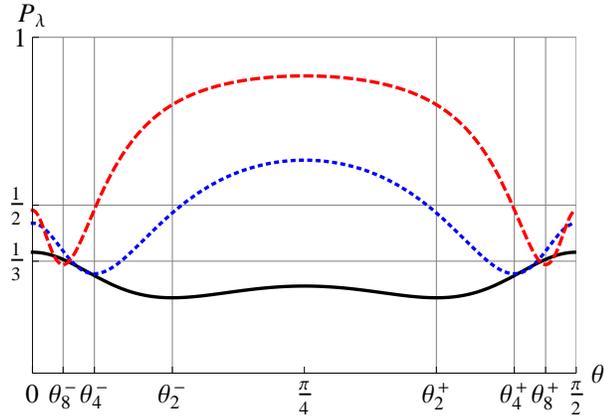}
\caption{ \label{purity1theta} (Color online) Purity $P_\lambda$, for  $\lambda=1$,  as a function of the polar angle
 $\theta=\arctan(r_-/r_+)$ for $r=2$ (solid line) $r=4$ (dotted blue) and $r=8$ (dashed red). The points $\theta^\pm_r$ denote
 local minima of $P_\lambda$ for each value of $r$ and are marked with vertical grid lines. Horizontal grid lines denote limit
 $r\to\infty$ values: $P_\lambda\to 1$  for $\theta\not=\theta^\pm_r, 0,\pi/2$; $P_\lambda\to 1/2$ for  $\theta=0,\pi/2$;
 $P_\lambda\to 1/3$ for $\theta=\theta^\pm_r$.}
\end{figure}

Using Wigner matrix properties like
\[\sum_{q=-j}^j\cD^{j}_{q,q}(X)=\!\!\!\!\!\sum_{h=\frac{\mathrm{odd}[2j]}{2}}^j(-1)^{j-h}\binom{j+h}{2h}\det(X)^{j-h}\tr(X)^{2h},\]
[with $\mathrm{odd}[n]=((-1)^{n+1}+1)/2$], purity \eqref{purityb} can be written only in terms of the $U(2)^2$ invariants (trace and determinant) as
\be
\tr(\varrho_b^2)=\sum_{n=0}^\lambda\sum_{k=0}^n C_{n,k}^\lambda\frac{\det(Z^\dag Z)^{n-k}
\tr(Z^\dag Z)^{2k}}{\det(\sigma_0+Z^\dag Z)^{2\lambda}},
\ee
with $C_{n,k}^\lambda$ certain coefficients(we do not give their cumbersome expression), which reproduce \eqref{purityl1} for $\lambda=1$. 
Adopting the decomposition \eqref{decompZ} of a matrix $Z$,  the CS purity $P_\lambda=\tr(\varrho_b^2)$ for general $\lambda$
can be written as a function of $r_+$ and $r_-$ of the form
\be
P_\lambda(r_+,r_-)=\sum_{n=0}^{2\lambda} f_n(r_+,r_-)\frac{(r_+r_-)^{2n}}{\left((1+r_+^2)(1+r_-^2)\right)^{2\lambda}},
\ee
with
\[f_n(r_+,r_-)=\!\!\!\sum_{j=\frac{\mathrm{odd}[n]}{2}}^{\lambda-\frac{n}{2}}\!\!\!\frac{\binom{\lambda+1}{\frac{n}{2}+j+1}
\binom{\lambda+1}{\frac{n}{2}-j}}{\lambda+1}\frac{\left(\frac{r_+}{r_-}\right)^{4j}\!\!\!r_+^4-
\left(\frac{r_-}{r_+}\right)^{4j}\!\!\!r_-^4}{r_+^4-r_-^4}.\]
Purity has the following invariant inversion property
\be
P_\lambda(r_+,r_-)=P_\lambda(\frac{1}{r_\mp},\frac{1}{r_\pm}).\label{purinv}
\ee
Figure \ref{purity13d} represents the CS purity $P_\lambda$ and its standard deviation $\Delta P_\lambda$ for $\lambda=1$ as a function
of $r_\pm$.
Purity is minimum at $r_\pm=1$ (maximum interlayer entanglement). One can also see that there is no interlayer entanglement 
(purity $P_\lambda=1$) for $r_\pm=0$ (which means all flux quanta in layer $b$)   
and when $r_\pm\to\infty$  (which means all flux quanta in layer $a$), except when $r_+\cdot r_-=0$,  for
which purity tends to $P_\lambda=1/(\lambda+1)$ when $r_\pm\to\infty$. There are other two particular trajectories in the $r_\pm$ plane for which
there is always interlayer entanglement.
To better appreciate this fact, we also represent in Figure \ref{purity13dpol} the purity  and its standard deviation for $\lambda=1$ as a function
of $r=\sqrt{r_+^2+r_-^2}$ and $\theta=\arctan(r_-/r_+)$.
For $r> 1.55$ and $\lambda=1$ the purity $P_\lambda$ displays two local minima (for fixed $r$) at two values
of the polar angle $\theta^\pm_r$ given by
\be
\cos\theta^\pm_r=\sqrt{\frac{3r^2+2r^4\mp\sqrt{-36-36r^2-11r^4+4r^6+4r^8}}{6r^2+4r^4}}.\label{thetapm2}
\ee
The expression  $\theta_\pm(r)=\theta^\pm_r$ gives
two singular trajectories in the $(r,\theta)$ plane for which the CS $|Z\ra$ remains always entangled. In fact,
purity tends to $P_1=1/3$ when $r\to\infty$ on these two trajectories.
Both local minima are narrower and narrower, with
$\theta^-_r\to 0$ and $\theta^+_r\to\pi/2$ when $r\to\infty$.
Purity fluctuations $\Delta P_\lambda$  are also high around these
two trajectories, as can be appreciated in Figure \ref{purity13dpol} (bottom panel).
See Figure \ref{purity1theta} for a plot of three sections,
$r=2, r=4$ and $r=8$, of $P_1$ as a function of $\theta$. The situation here is similar to the one depicted in Figure \ref{imbalancedevtheta}
for the imbalance standard deviation.

Let us also examine the particular (diagonal) case $r_+=r_-=r_0$, for which purity simplifies to
\be
P_\lambda(r_0)=\frac{\sum_{n=0}^{2\lambda}\binom{\lambda}{\frac{n+\mathrm{odd}[n]}{2}}
\binom{\lambda}{\frac{n-\mathrm{odd}[n]}{2}}r^{4n}_0}{(1+r^2_0)^{4\lambda}}.\label{purityspinnig}
\ee
In Figure \ref{purity} we represent  purity $P_\lambda$ and its fluctuations $\Delta P_\lambda$ as a function of $r_0$ for different values of $\lambda$.
We see that purity $P_\lambda$ of a CS $|Z\rangle$ is minimum
(maximum interlayer entanglement) at $r_0=1$ for all values of $\lambda$ (the vertical grid line indicates this 
particular value of $r_0$ for which maximum interlayer entanglement is attained). 
Actually, as already noticed in eq. \eqref{purinv}, purity is invariant under inversion
$P_\lambda(r_0)=P_\lambda({1}/{r_0})$, with $r_0=1$ a fixed point.
However, the CS $|Z\rangle$ is never 
maximally entangled since $P_\lambda(1)$ is always greater than $1/d_\lambda$ (purity of a completely mixed state). 
In particular, for $\lambda=1$ we have 
$P_1(1)=3/16$ (see Figure \ref{purity}), which is slightly greater than $1/d_1=1/6$. 
Purity fluctuations also display a local minimum at $r_0=1$ (see Figure \ref{purity}), which becomes flatter and flatter 
as $\lambda$ increases. We also appreciate that interlayer entanglement of $|Z\rangle$ attains its maximum (zero purity) in a wide
neighborhood of $r_0=1$ for high values of $\lambda$, this making entanglement robust under perturbations (purity
fluctuations are also negligible in this limit in the region around $r_0=1$). The horizontal grid line indicates
the pure-state purity, which is attained at $r_0=0$ [all particles in layer $b$, with state \eqref{z0}]  and when $r_0\to\infty$  
[all particles in layer $a$, with state \eqref{zinf}].

Now we shall compare the particular case $r_+=r_-=r_0$ with the spin-frozen case with pseudospin-$s$. 
We see that the purity $p_s$ in eq. \eqref{puritysu2eq} for the spin-frozen case does not coincide with the purity $P_\lambda$ 
for the spinning case in eq. \eqref{purityspinnig}, although Figure \ref{puritysu2} displays a similar qualitative behavior of $p_s(r)$ 
with respect to $P_\lambda(r)$ in Figure \ref{purity} (we must compare $2s\leftrightarrow\lambda$, the total number of magnetic flux lines piercing 
one electron). 
The difference between $P_\lambda(r)$ and $p_s(r)$ indicates that spin degrees of freedom play a role in the interlayer entanglement by, for example,
making it more robust than in the spin-frozen case, as commented before. Indeed, on the one hand, maximum interlayer entanglement (zero purity) 
is attained for high values of $\lambda$ in a wide interval of the tunneling interaction strength around $r_0=1$. On the other hand, purity 
fluctuations are also negligible inside this tunneling interaction strength region.

\begin{figure}
\includegraphics[width=8cm]{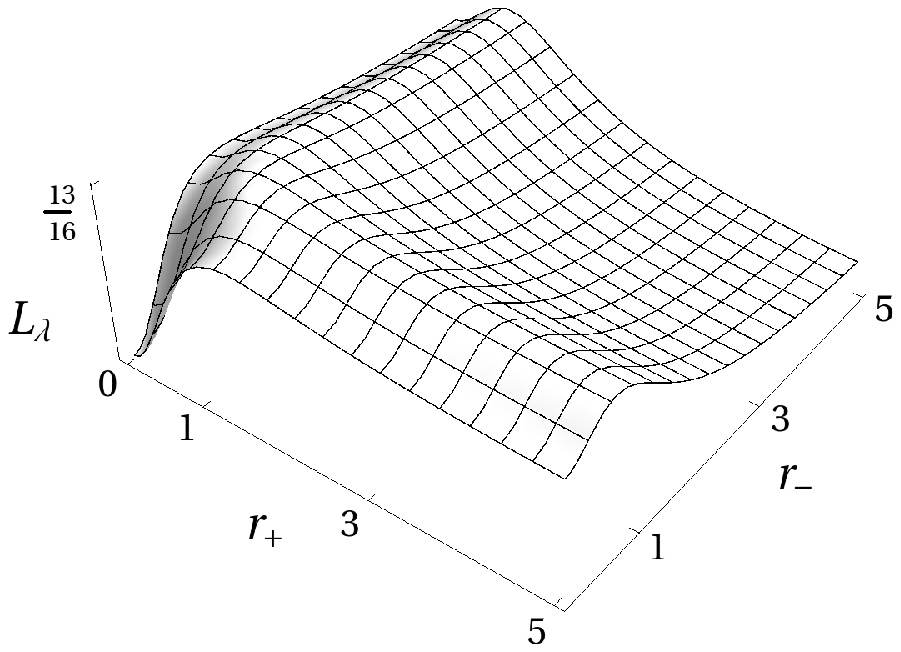}
\includegraphics[width=8cm]{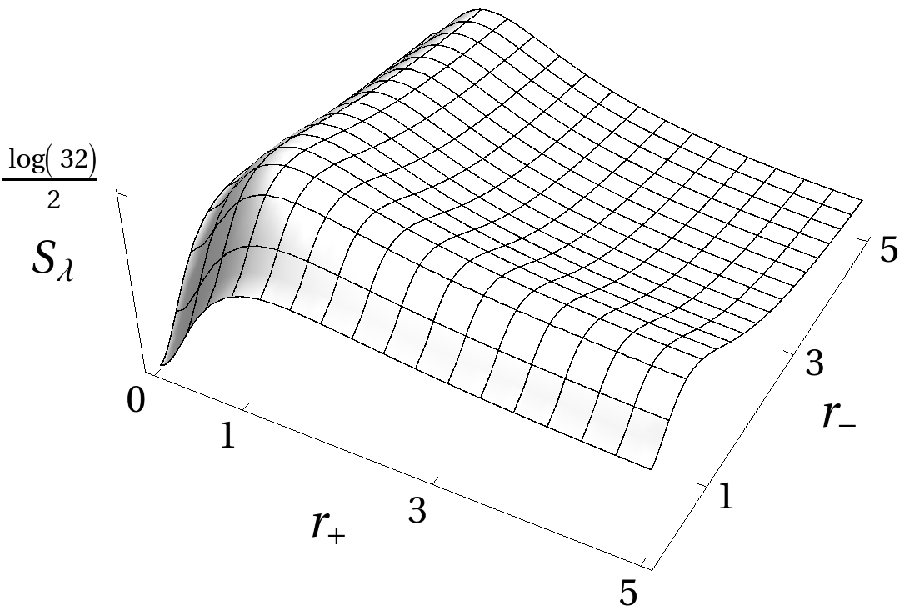}
\caption{ \label{LinVon3D} Linear $L_\lambda=1-P_\lambda$ and  Von Neumann $S_\lambda$ entanglement 
entropies of the reduced density matrix $\varrho_b=\tr_a(\varrho)$
to layer $b$, for the CS density matrix $\varrho=|Z\rangle\langle Z|$, as a function of the CS parameters
$r_\pm$ for $\lambda=1$.}
\end{figure}

\begin{figure}
\includegraphics[width=8cm]{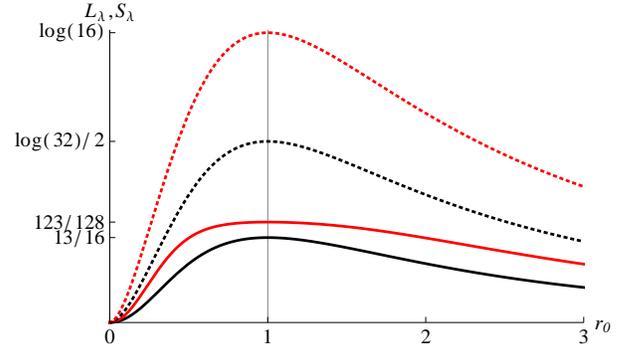}
\caption{ \label{LinVon} (Color online) Linear $L_\lambda$ (solid line) and Von Neumann $S_\lambda$ (dotted line) 
entanglement entropies of the reduced density matrix $\varrho_b=\tr_a(\varrho)$
to layer $b$, for the CS density matrix $\varrho=|Z\rangle\langle Z|$, as a function of the CS parameter
$r_0\equiv r_+=r_-$  for two values of: $\lambda=1$ (black) and $\lambda=2$ (red).}
\end{figure}

For those readers who prefer Von Neumann to linear entanglement entropy, it is also possible to compute 
$S_\lambda(r_+,r_-)=-\tr(\varrho_b\log\varrho_b)$ for $\varrho_b$ in \eqref{rdmcs}. Taking into account that $\varrho_b$ is block-diagonal and 
after a little bit of algebra, we arrive to the following expression for 
\bea
S_\lambda(r_+,r_-)&=&-\sum_{m=0}^\lambda \sum_{j=0;\um}^{(\lambda-m)/2}\sum^{j}_{q=-j}
(2j+1) \\ &&\times\gamma^{m}_{j,q}(r_+,r_-)\log \gamma^{m}_{j,q}(r_+,r_-),\nn
\eea
with 
\be
\gamma^{m}_{j,q}(r_+,r_-)=\frac{\binom{\lambda+1}{\lambda-2j-m}
\binom{\lambda+1}{\lambda-m+1}}{\lambda+1}\frac{r_+^{2(j+m+q)}r_-^{2(j+m-q)}}{(1+r_+^2)^\lambda (1+r_-^2)^\lambda}.\nn
\ee
In Figure \ref{LinVon3D} we perceive a similar qualitative behavior of linear $L_\lambda=1-P_\lambda$ 
and Von Neumann $S_\lambda$ entropies for $\lambda=1$. For general $\lambda$ the situation is similar, with $L_\lambda$ a lower 
approximation of $S_\lambda$.
In Figure \ref{LinVon} we plot $L_\lambda$ and $S_\lambda$ as a function of $r_0\equiv r_+=r_-$. We see that 
$S_\lambda\geq L_\lambda$ and that the maximum linear $L_\lambda^{\mathrm{max.}}=1-1/d_\lambda$ and Von Neumann 
$S_\lambda^{\mathrm{max.}}=\log(d_\lambda)$ entropies are never attained, although $|Z\rangle$ is almost 
maximally entangled for $r_0=1$, where $L_1(1)=13/16 \lesssim L_1^{\mathrm{max.}}=5/6$ and 
$S_1(1)=\log(\sqrt{32}) \lesssim S_1^{\mathrm{max.}}=\log(6)$ (see maximum values for $\lambda=1$ and $\lambda=2$ in Figure \ref{LinVon}).

\subsection{Interlayer entanglement of a thermal state}

In the two previous cases we have studied the interlayer entanglement of pure bipartite states. In this Section we tackle 
the study of a mixed state like the equilibrium state in BLQH system at
finite temperature. Other studies about entanglement spectrum and entanglement thermodynamics of BLQH systems 
at $\nu=1$ can be found in \cite{Schliemann}.

\begin{figure}
\includegraphics[width=8cm]{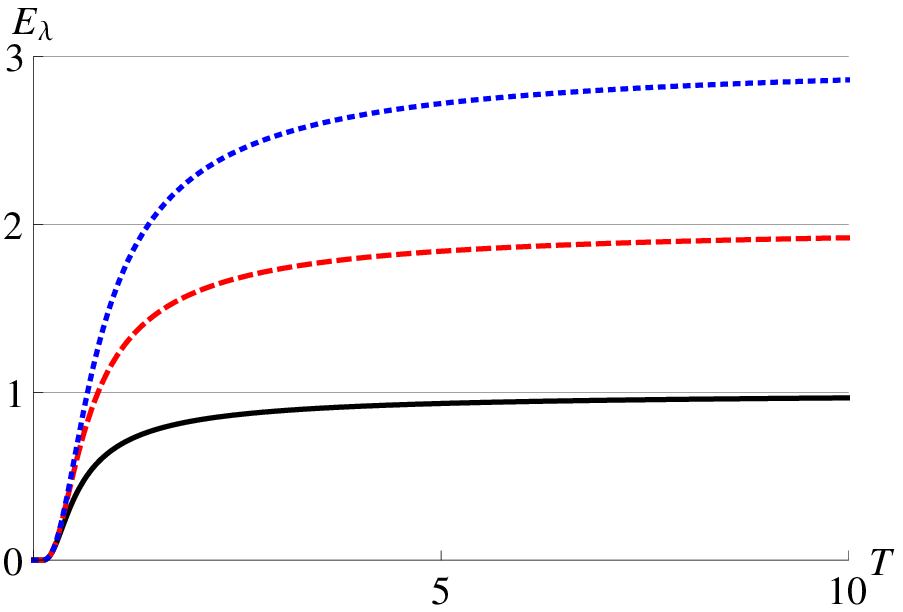}
\includegraphics[width=8cm]{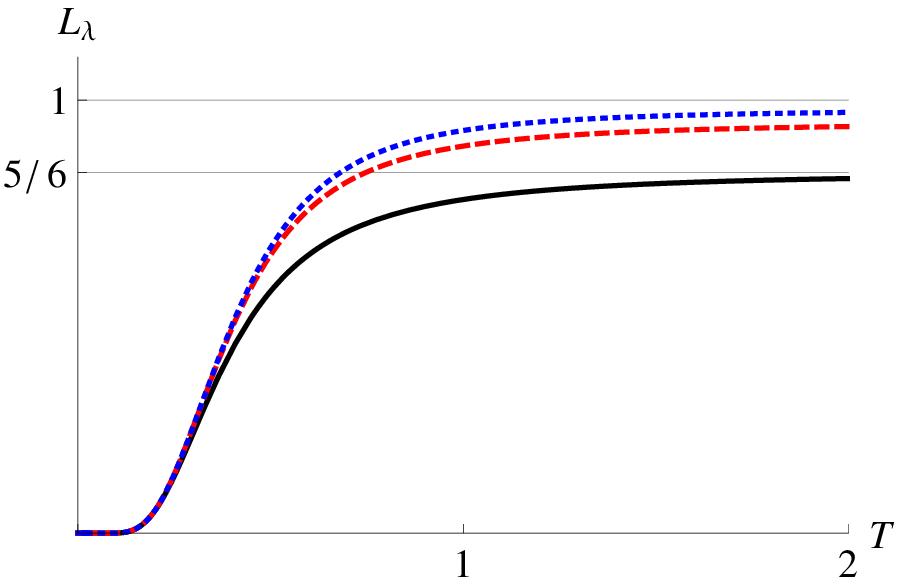}
\includegraphics[width=8cm]{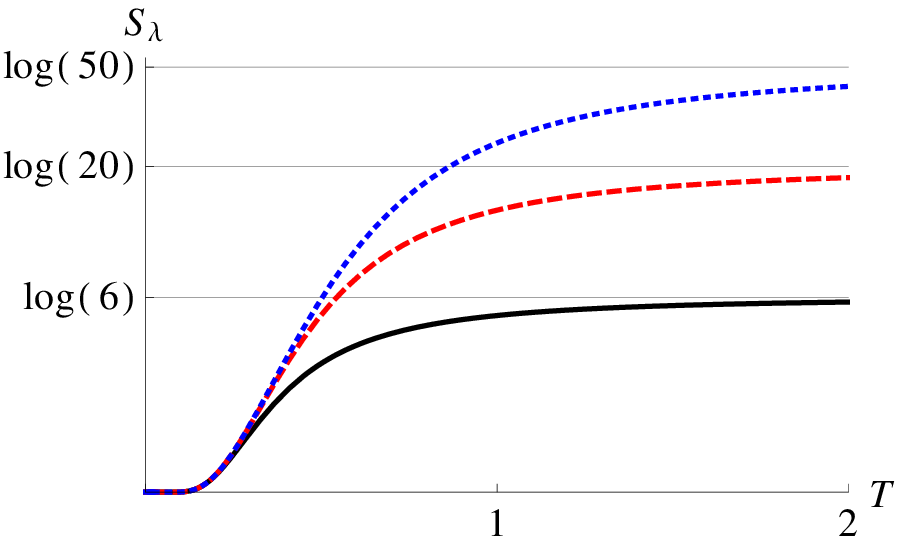}
\caption{ \label{EnVonBath} Mean energy $E_\lambda$, linear $L_\lambda$ and  Von Neumann  $S_\lambda$ entanglement 
entropy of a thermal equilibrium state $\rho_\lambda$ as a function of the temperature $T$ 
for $\lambda=1$ (black), $\lambda=2$ (dashed red) and $\lambda=3$ (dotted blue). We are taking $\varepsilon, k_B=1$ unities.}
\end{figure}

\begin{figure}
\includegraphics[width=8cm]{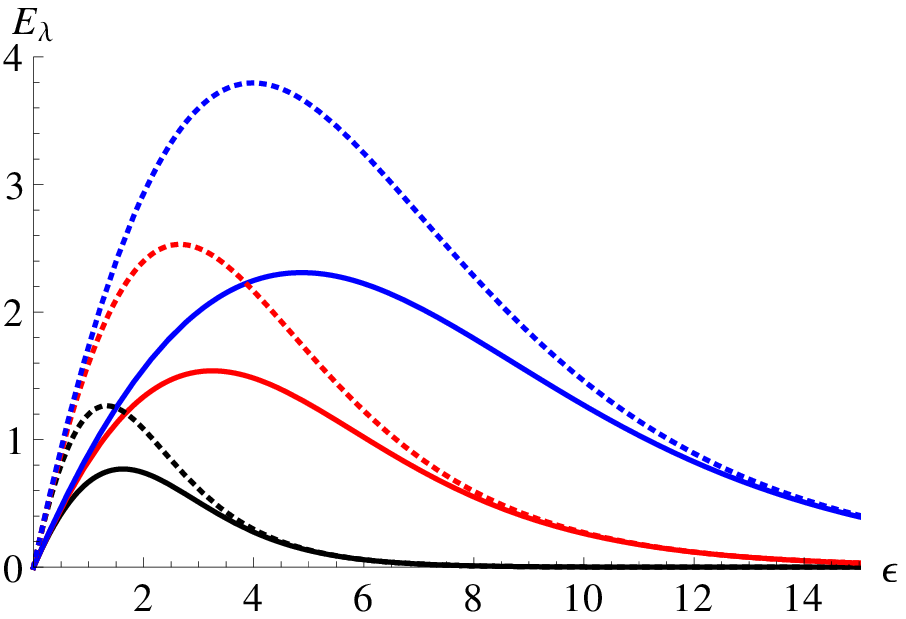}
\includegraphics[width=8cm]{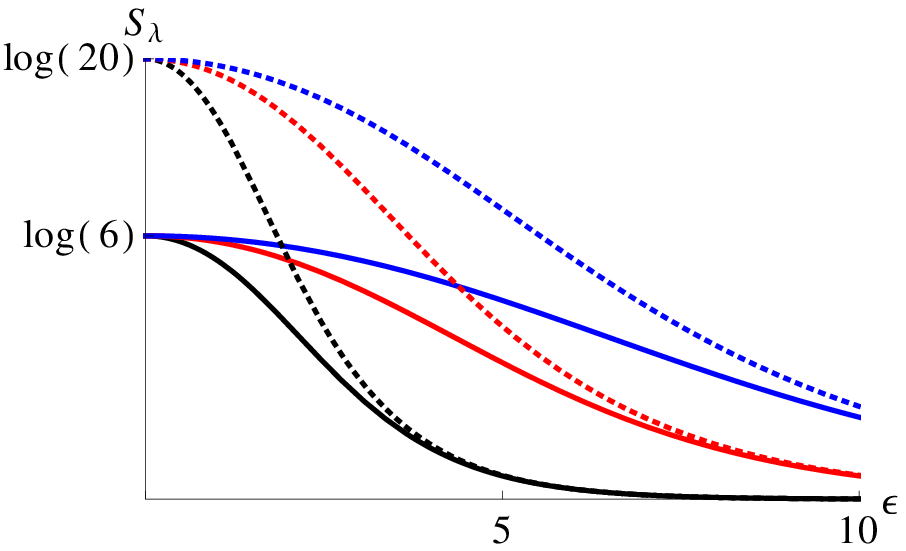}
\caption{ \label{EnBias} Mean energy $E_\lambda$ and Von Neumann entropy $S_\lambda$ of a thermal equilibrium state $\rho_\lambda$ as a function of the bias 
voltage $\varepsilon$ 
for $\lambda=1$ (solid) and  $\lambda=2$ (dotted) and several temperatures: $T=1$ (black), $T=2$ (red) and $T=3$ (blue). We are taking $k_B=1$ unities.}
\end{figure}

For the sake of simplicity, we shall consider the pseudo-Zeeman Hamiltonian given by 
$H=\varepsilon(\mathcal P_3+\lambda)$, where $\varepsilon$ is the bias voltage parameter (it introduces an energy scale into the system) 
and we have added a zero-point energy $\varepsilon\lambda$ for convenience. The basis states $|{}{}_{q_a,q_b}^{j,m}\ra$ are 
Hamiltonian eigenvectors with eigenenergies $\mathcal{E}_{n}=\varepsilon n$, with $n=2j+2m$. The degeneracy $D_n$ of the energy level $\mathcal{E}_n$  
depends on $n=2j+2m$ in the form:
\be
D_n=\left\{\ba{l} \frac{(n+1)(n+2)(n+3)}{6}, \,n\leq\lambda,\\ \frac{(2\lambda-n+1)(2\lambda-n+2)(2\lambda-n+3)}{6},\, 
\lambda\leq n\leq 2\lambda.\ea\right.
\ee
Formula \eqref{dimensionj} can be alternatively written as $\sum_{n=0}^{2\lambda}D_n={d_\lambda}$ in terms of the degeneracy $D_n$. 
The normalized density matrix is written in compact form as $\rho_\lambda(\beta)=e^{-\beta H}/\tr(e^{-\beta H})$, with $\beta=1/(k_BT)$ ($k_B$ denotes the 
Boltzmann constant and $T$ the temperature), as usual. 
The canonical partition function is easily calculated and gives
\bea
Q_\lambda(\beta)&=&\sum_{n=0}^{2\lambda} D_n e^{-\beta \mathcal{E}_n}\\ 
&=&\frac{1+e^{-\beta\varepsilon(2\lambda+4)}+2(\lambda+1)(\lambda+3)e^{-\beta\varepsilon(\lambda+2)}
}{(1-e^{-\beta\varepsilon})^4}\nonumber\\ 
&& -(2+\lambda)^2\frac{e^{-\beta\varepsilon(\lambda+1)}+e^{-\beta\varepsilon(\lambda+3)}
}{(1-e^{-\beta\varepsilon})^4}.\nonumber
\eea
One can check that at high temperatures $\lim_{\beta\to 0}Q_\lambda(\beta)=d_\lambda$ (the dimension of the Hilbert space). 
The mean energy can be calculated either directly as $E_\lambda(\beta)=\sum_{n=0}^{2\lambda} D_n \gamma_{n}(\beta)\mathcal{E}_n$,  
with $\gamma_{n}(\beta)=e^{-\beta \mathcal{E}_{n}}/Q_\lambda(\beta)$ the Boltzmann factor, or through the well known formula 
$E_\lambda(\beta)=-\partial\log(Q_\lambda(\beta))/\partial\beta$. In particular, we see that the mean energy at high 
temperatures is $E_\lambda(0)=\lambda\varepsilon$, and at zero temperature is $E_\lambda(\infty)=0$. In Figure \ref{EnVonBath} 
(top panel) we plot the mean energy as a function of the temperature $T=1/(k_B\beta)$ in $\varepsilon=1$ unities.
It is also interesting to see the representation of the mean energy $E_\lambda$ as a function of the bias voltage $\varepsilon$ in Figure 
\ref{EnBias} (top panel), where one can observe a similarity with the energy of a black body as a function of the frequency $\omega$; In fact, 
the spectrum is peaked at a characteristic bias voltage $\varepsilon_c$ (resp. frequency $\omega_c$) that shifts to higher voltages 
(resp. frequencies) with increasing temperature; this reminds the Wien's displacement law $\beta\varepsilon=c(\lambda)$, with $c(\lambda)$ 
a ``Wien's displacement constant'' depending on $\lambda$. Note that here we have an extra 
parameter $\lambda$ to play with.

The entropy $S_\lambda(\beta)=-\tr[\rho_\lambda(\beta)\log\rho_\lambda(\beta)]$ can be calculated either directly as
the formula 
\be
S_\lambda(\beta)=-\sum_{n=0}^{2\lambda} D_n \gamma_{n}(\beta) 
\log \gamma_{n}(\beta),
\ee
or through the general formula
\be
S_\lambda(\beta)=\beta E_\lambda(\beta)+\log Q_\lambda(\beta).
\ee

The reduced density matrix to layer $b$,  $\rho_\lambda^b=\tr_a(\rho_\lambda)$, is
\be
 \rho_\lambda^b=\sum_{m=0}^\lambda \sum_{j=0;\um}^{(\lambda-m)/2}\sum^{j}_{q,q_b=-j}\gamma_{n}(\beta)
 |v_{q,q_b}^{j,\lambda-2j-m}\ra_b\la v_{q,q_b}^{j,\lambda-2j-m}|,
\ee
($n=2j+2m$) whose purity is easily calculated in terms of the partition function as
\be
P_\lambda(\beta)=\sum_{n=0}^{2\lambda} D_ n\gamma_{n}(\beta)^2=\frac{Q_\lambda(2\beta)}{Q_\lambda(\beta)^2}.
\ee
In Figure \ref{EnVonBath}, middle  panel, we represent the linear entropy $L_\lambda=1-P_\lambda$ as a function of the temperature 
for three values of $\lambda$. We see that $L_\lambda$ is 
zero at zero temperature and $L_\lambda\to 1-1/d_\lambda$ at high temperatures, where the state is maximally entangled. In the same way, we can 
compute the Von Neumann entanglement 
entropy $S_\lambda^b=-\tr[\rho_\lambda^b\log\rho_\lambda^b]$ which,  after a little bit of algebra, we arrive to the conclusion that $S_\lambda^b(\beta)=
S_\lambda^a(\beta)=S_\lambda(\beta)$; that is, the entropy restricted to any of the layers coincides 
with the total bilayer entropy. In particular, the subadditivity condition $S_\lambda\leq 
S_\lambda^a+S_\lambda^b=2S_\lambda$ is fulfilled. In Figure \ref{EnVonBath} (bottom panel) we 
plot the entropy $S_\lambda$ as a function of $T$ for three values of $\lambda$. We see that $S_\lambda$ is 
zero at zero temperature and $S_\lambda\to\log(d_\lambda)$ at high temperatures, where the state is maximally entangled, in accordance 
with the results of the linear entropy $L_\lambda$ (which is a lower bound of $S_\lambda$).  We also represent $S_\lambda$ as a function of the 
bias voltage $\epsilon$ in Figure \ref{EnBias} (bottom panel). We see that $S_\lambda=\log(d_\lambda)$ (maximal) at zero bias voltage $\varepsilon=0$ and 
goes to zero for high $\varepsilon$. Entropy $S_\lambda$ grows with $\lambda$ and $T$ for fixed $\epsilon$.

\section{Conclusions and outlook}\label{comments}

We have studied interlayer imbalance and entanglement (and its fluctuations) of basis states $|{}{}_{q_a,q_b}^{j,m}\ra$, coherent states $|Z\ra$, 
and mixed thermal states $\rho_\lambda$ in the state space (at one Landau site) of the BLQH system at filling factor $\nu=2/\lambda$. 
Isospin-$\lambda$ CS are labeled by $2\times 2$ complex matrices $Z$  in the 8-dimensional Grassmannian manifold $\mathbb{G}_2=U(4)/U(2)^2$ and generalize 
the standard pseudospin-$s$ CS $|z\ra$ labeled by complex points $z\in \mathbb{S}^2=U(2)/U(1)^2$ (the Riemann-Bloch sphere). 
The interplay between spin and pseudospin (layer) degrees of freedom introduces novel physics with regard to the spin-frozen case, 
by making interlayer entanglement more robust for a wide range of coherent state parameters 
(specially for high values of the number $\lambda$ of magnetic flux lines). Von Neumann 
entanglement entropy of mixed thermal states is maximal at high temperatures and zero bias voltage (when we consider a pseudo-Zeeman Hamiltonian). 

Other bipartite entangled BLQH systems (namely, spin-pseudospin \cite{Yusuke} or electron-electron) might also be considered which could also be of interest 
in quantum information theory. We must say that entangled (usually oscillator and spin) coherent states are important to quantum superselection principles, 
quantum information processing and quantum optics, where they have been produced in a conditional propagating-wave 
realization (see e.g. \cite{entangcoh-review} for a recent review on the subject). Coherent (quasi-classical) states are easily generated for 
many interesting physical systems, and we believe that BLQH CS $|Z\rangle$ at $\nu=2/\lambda$ will not be exception and that they will play 
an important role, not only in theoretical considerations but, also in experimental settings. 
Another interesting possibility is to study entanglement between two different spatial regions. Before, we should extend the present study 
to several Landau sites. In this case, the Coulomb exchange Hamiltonian, which is described by an anisotropic $U(4)$
nonlinear sigma model  in BLQH systems, provides the necessary interaction to create quantum correlations between spatial regions.

Other $U(4)$ operator mean values $\la Z|\mathcal{T}_{\mu\nu}|Z\ra$ (and their powers) can also be calculated, which could be specially suitable to
analyze the classical (thermodynamical or mean-field limit) and phase diagrams of BLQH Hamiltonian models 
undergoing a quantum phase transition, like the well studied spin-ferromagnet
and pseudo-spin-ferromagnet phases at $\nu=1$, or the spin, ppin and canted phases at $\nu=2$. 
Actually, coherent states for other symmetry groups [viz, Heisenberg, $U(2)$ and $U(3)$] already provided 
essential information about the quantum phase transition occurring in several interesting models like for example: 
the Dicke model for atom-field ineractions  \cite{casta2,epl2012,renyipra,husidi}), vibron models for molecules 
\cite{curro,husivi}, and also pairing models like the Lipkin-Meshkov-Glick model for 
nuclei \cite{LMG,pairons}. For vibron models, $U(3)$ coherent states have been used as variational 
states capturing rovibrational entanglement of the ground state in shape phase transitions of 
molecular benders \cite{entangvib,entangvib2}. We also believe that the proposed Grassmannian coherent states $|Z\rangle$ 
can provide valuable physical information about the ground state and phase diagram in the semi-classical limit of BLQH systems at $\nu=2/\lambda$. 
This is work in progress. 

To conclude,  we would like to mention that graphene physics shares similarities with BLQH systems, where the 
two valleys (or Dirac points) play a role similar to the layer degree of freedom. 
Other Grassmannian cosets $U(N)/[U(M)\times U(N-M)]$ appear in this context (see e.g. \cite{KunYang3}) and we 
believe that a boson realization like the one discussed here can contribute something interesting 
also in this field.

\section*{Acknowledgements}

Work partially supported by the
Spanish MINECO, Junta de Andaluc\'\i a and University of Granada under projects FIS2011-29813-C02-01, 
FQM1861 and [CEI-BioTIC-PV8 and PP2012-PI04], respectively. 

\appendix

\section{Orthonormal basis for arbitrary $\lambda$}\label{basissubsec}

In Ref. \cite{GrassCSBLQH} we have generalized, in a natural way, the Fock space realization of pseudospin-$s$ basis states 
\eqref{basisinfocksu2} to a Fock space representation of the basis functions $|{}{}_{q_a,q_b}^{j,m}\ra$ of
${\cal H}_\lambda(\mathbb G_2)$. We have found a $U(4)$ generalization of the $U(2)$ monomials $\varphi_k(z)$ in eq. \eqref{basisinfocksu2} and \eqref{su2cs} 
in terms of a set of homogeneous polynomials of degree $2j+2m$
\bea
\varphi_{q_a,q_b}^{j,m}(Z)&=&\sqrt{\frac{2j+1}{\lambda+1}\binom{\lambda+1}{2j+m+1}\binom{\lambda+1}{m}}\label{basisfunc}\\
&&\times \det(Z)^{m}\cD^{j}_{q_a,q_b}(Z),\; \begin{matrix}
2j+m\leq\lambda, \\ q_a,q_b=-j,\dots,j, \end{matrix}\nn\eea
in four complex variables $z^\mu=\tr(Z\sigma_\mu)/2, \mu=0,1,2,3$, where 
\bea
&& \cD^{j}_{q_a,q_b}(Z)=\sqrt{\frac{(j+q_a)!(j-q_a)!}{(j+q_b)!(j-q_b)!}}
 \sum_{k=\max(0,q_a+q_b)}^{\min(j+q_a,j+q_b)}\label{Wignerf}\\ 
&& \binom{j+q_b}{k}\binom{j-q_b}{k-q_a-q_b}   z_{11}^k
z_{12}^{j+q_a-k}z_{21}^{j+q_b-k}z_{22}^{k-q_a-q_b},\nn\eea
denotes the usual Wigner $\cD$-matrix \cite{Louck3} 
for a general $2\times 2$ complex matrix $Z$ with entries $z_{jk}$ and angular momentum $j$. The set  \eqref{basisfunc} verifies the
closure relation
\be\sum^{\lambda}_{m=0}\!\!\sum_{j=0;\um}^{(\lambda-m)/2}\!\!\sum^{j}_{q_a,q_b=-j}\!\!
\overline{\varphi_{q_a,q_b}^{j,m}({Z'})}\varphi_{q_a,q_b}^{j,m}(Z)={\det(\sigma_0+Z'^\dag
Z)^\lambda}\nn\ee
which is the $U(4)$ version of the more familiar $U(2)$ closure relation $\sum_{k=-s}^s\overline{\varphi_k(z')}\varphi_k(z)=(1+\bar z'z)^{2s}$ 
leading to the pseudospin-$s$ CS overlap \eqref{su2overlap}.

With this information, and treating $\varphi_{q_a,q_b}^{j,m}$ as polynomial creation and annihilation operator functions [like the monomials $\varphi_k$ in 
\eqref{basisinfocksu2}], we have found in Ref. \cite{GrassCSBLQH}
that the set of orthonormal basis vectors \eqref{basisvec} can be obtained in terms of Fock states \eqref{grassmannbasis2} as
\bea
|{}{}_{q_a,q_b}^{j,m}\ra&=&\frac{1}{\sqrt{2j+1}}\sum_{q=-j}^{j}(-1)^{q_a-q}\label{basisinfock2}\\
&\times&\frac{\varphi^{j,m}_{-q,-q_a}(\mathbf{a}^\dag)}{\sqrt{\frac{\lambda!(\lambda+1)!}{(\lambda-2j-m)!(\lambda+1-m)!}}}
\frac{\varphi^{j,\lambda-2j-m}_{q,q_b}(\mathbf{b}^\dag)}{\sqrt{\frac{\lambda!(\lambda+1)!}{m!(2j+m+1)!}}}
\;|0\ra.\nn
\eea
This is the $U(4)$ version of eq. \eqref{basisinfocksu2} for the pseudospin-$s$ basis states $|k\ra$ of $U(2)$,
with the role of $s$ played now by $\lambda$. However, as we proof in Subsect. \ref{entbasissec}, whereas
the state $|k\ra$ is a direct product and does not entangle layers  $a$ and $b$, the state
 $|{}{}_{q_a,q_b}^{j,m}\ra$ does entangle both layers
for angular momentum $j\not=0$. This is better seen when we define the set of (Schmidt) states for layer $a$ (idem for layer $b$)
\be
|v_{q,q'}^{j,m}\ra_a=\frac{\varphi^{j,m}_{q,q'}(\mathbf{a}^\dag)}{\sqrt{\frac{\lambda!(\lambda+1)!}{(\lambda-2j-m)!(\lambda+1-m)!}}}|0\ra,
\label{basislayer}
\ee
and realize that it constitutes an orthonormal set for this layer, that is
\be
\la v_{p,q}^{j,m}|v_{p',q'}^{j',m'}\ra_a=\delta_{j,j'}\delta_{m,m'}\delta_{p,p'}\delta_{q,q'}.
\label{ortholayer}
\ee
With this notation, the expression \eqref{basisinfock2} becomes \eqref{basisinfock1}.

Concerning the quantum statistics of our states for a given number $\lambda$ of flux lines, we have already 
mentioned in eq. \eqref{statistics} that the basis states $|{}{}_{q_a,q_b}^{j,m}\ra$ 
 are antisymmetric (fermionic character) under the 
interchange of the two electrons for $\lambda$ odd, and they are symmetric (bosonic character) for $\lambda$ even. Indeed, under the 
interchange of columns in \eqref{atilde}  the operator functions \eqref{basisfunc} verify 
$\varphi^{j,m}_{q_a,q_b}(\widetilde{\mathbf{a}}^\dag)=(-1)^{m}\varphi^{j,m}_{-q_a,q_b}(\mathbf{a}^\dag)$. Taking into account 
that $(-1)^{2q}=(-1)^{2j}$ for any $q=-j,\dots,j$ and doing some algebraic manipulations, one arrives to the identity 
\eqref{statistics}, where the left-hand side vector is constructed as in \eqref{basisinfock2} 
but replacing $\mathbf{a}^\dag$ and $\mathbf{b}^\dag$ by $\widetilde{\mathbf{a}}^\dag$ and $\widetilde{\mathbf{b}}^\dag$, 
respectively, that is, switching both electrons.  

\section{Coherent states for arbitrary $\lambda$}\label{subsecCS}

The extension of the formula \eqref{cs1} (for $\lambda=1$) to arbitrary $\lambda$ has been worked out in Ref. \cite{GrassCSBLQH}. 
Here we reproduce it for the sake of self-containedness. CS $|Z\rangle$ 
 are labeled by a $2\times 2$ complex matrix $Z=z^\mu\sigma_\mu$ (sum on $\mu=0,1,2,3$), with four
complex coordinates  $z^\mu=\tr(Z\sigma_\mu)/2$,  and  can be expanded in terms of
the orthonormal basis vectors \eqref{basisinfock2} as
\be
|Z\ra=\frac{\sum^{\lambda}_{m=0}\sum_{j=0;\um}^{(\lambda-m)/2}\sum^{j}_{q_a,q_b=-j}\varphi_{q_a,q_b}^{j,m}(Z)
|{}{}_{q_a,q_b}^{j,m}\ra}{\det(\sigma_0+Z^\dag Z)^{\lambda/2}},\label{u4cs}
\ee
with coefficients $\varphi_{q_a,q_b}^{j,m}(Z)$ in \eqref{basisfunc}. Denoting by $\check{\mathbf{a}}=\um\eta^{\mu\nu}\tr(\sigma_\mu\mathbf{a})\sigma_\nu$ and
$\check{\mathbf{b}}=\um\eta^{\mu\nu}\tr(\sigma_\mu\mathbf{b})\sigma_\nu$ [we are using Einstein summation convention with Minkowskian
metric $\eta_{\mu\nu}=\mathrm{diag}(1,-1,-1,-1)$] the ``parity reversed'' $2\times 2$-matrix annihilation operators 
of $\mathbf{a}$ and $\mathbf{b}$, the CS $|Z\ra$ in  \eqref{u4cs} can also be written in the form of a boson condensate as
\be
|Z\ra=\frac{1}{\lambda!\sqrt{\lambda+1}}\left(\frac{\det(\check{\mathbf{b}}^\dag+
Z^t\check{\mathbf{a}}^\dag)}{\sqrt{\det(\sigma_0+Z^\dag Z)}}\right)^\lambda|0\ra,
\label{u4csfock}
\ee
with $|0\ra$ the Fock vacuum. All CS $|Z\rangle$, with  $Z\in \mathbb{G}_2$, are normalized, $\la Z|Z\ra=1$, 
but they do not constitute an orthogonal set since they have a non-zero (in general) overlap
given by
\be
\la Z'|Z\ra=\frac{\det(\sigma_0+Z'^\dag Z)^\lambda}{\det(\sigma_0+Z'^\dag Z')^{\lambda/2}\det(\sigma_0+Z^\dag Z)^{\lambda/2}}\label{u4csov}
\ee
However,  using  orthogonality properties of the homogeneous polynomials
$\varphi_{q_a,q_b}^{j,m}(Z)$, it is direct to prove that CS \eqref{u4cs} fulfill  the
resolution of unity
\be
1=\int_{\mathbb G_2} |Z\ra\la Z| d\mu(Z,Z^\dag),\label{u4csclos}
\ee
with $d\mu(Z,Z^\dag)=\frac{12d_\lambda}{\pi^4}\frac{\prod_{\mu=0}^3 d\Re(z^\mu) d\Im(z^\mu)}{\det(\sigma_0+Z^\dag Z)^{4}}$ the integration measure 
[this is the $U(4)$ generalization of the $U(2)$ integration measure on the sphere given after eq. \eqref{su2overlap}]. 
It is interesting to compare the $U(4)/U(2)^2$ CS in eqs. \eqref{u4cs} and \eqref{u4csfock} with the $U(2)/U(1)^2$ CS  in eqs. \eqref{su2cs} 
and \eqref{su2BE}, with CS overlaps \eqref{u4csov} 
and \eqref{su2overlap}, respectively. We perceive a similar structure between
$\mathbb{G}_2=U(4)/U(2)^2$ and $\mathbb{S}^2=U(2)/U(1)^2$ CS, although the Grassmannian $\mathbb{G}_2$ case is more
involved and constitutes a kind of ``matrix $Z$ generalization of the scalar $z$''.

For $Z=0$ we recover the lowest-weight state
\be |Z_0\ra=|{}{}_{q_a=0,q_b=0}^{j=0,m=0}\ra=\frac{\det(\mathbf b^\dag)^\lambda}{\lambda!\sqrt{\lambda+1}}|0\ra,\label{z0}\ee
with all $2\lambda$ flux quanta occupying the bottom layer $b$. For $Z\to\infty$ we recover the highest-weight state
\be |Z_\infty\ra=|{}{}_{q_a=0,q_b=0}^{j=0,m=\lambda}\ra=\frac{\det(\mathbf a^\dag)^\lambda}{\lambda!\sqrt{\lambda+1}}|0\ra,\label{zinf}\ee
with all $2\lambda$ flux quanta occupying the top layer $a$.

To finish, let us provide yet another expression of the CS $|Z\ra$ in  \eqref{u4cs}, now as an
exponential of interlayer ladder operators $\mathcal{T}_{\pm\mu}=(\mathcal{T}_{1\mu}\pm i\mathcal{T}_{2\mu})/2$ 
[remember their general definition \eqref{bosrepre}].
Let us denote by $\mathcal{T}_{+}\equiv
\mathcal{T}_{+\mu}\sigma^\mu=2\check{\mathbf{a}}^\dag\check{\mathbf{b}}$. The
CS $|Z\ra$ in  \eqref{u4cs} and \eqref{u4csfock} can also be written as the exponential action of the rising operators 
$\mathcal{T}_{+\mu}$ on the lowest-weight state $|{}{}_{q_a=0,q_b=0}^{j=0,m=0}\ra$ as 
\be
|Z\ra=\frac{e^{\um \tr(Z^t\mathcal{T}_{+})}}{\det(\sigma_0+Z^\dag Z)^{\lambda/2}}|{}{}_{0,0}^{0,0}\ra.\label{u4cs2}
\ee
This is the $U(4)$ version of the more familiar $U(2)$ (spin frozen) expression in eq. \eqref{su2cs}. In fact, for $\mu=0$ we have that
$\mathcal{T}_{+0}=(\mathcal{T}_{10}+ i\mathcal{T}_{20})/2=\mathcal{P}_1+i \mathcal{P}_2=\mathcal{P}_+$, according to the
usual notation in the literature \cite{EzawaBook} introduced in paragraph between equations \eqref{bosrepre} and  \eqref{grassmannbasis2}.

\end{document}